\pdfoutput=1
\documentclass[a4paper,
amsmath,amssymb,
twocolumn,
superscriptaddress,
10pt,
accepted=2019-04-23]{quantumarticle}

\usepackage[numbers, sort&compress]{natbib}
\usepackage[utf8]{inputenc}
\usepackage[english]{babel}
\usepackage[T1]{fontenc}
\usepackage{amsmath}
\usepackage{amsthm}

\usepackage{enumitem}
\usepackage{graphicx}
\usepackage{bm}
\usepackage{mathrsfs}
\usepackage{dsfont}
\usepackage{tikz}
\usetikzlibrary{decorations.pathmorphing}
\newcommand*{\f}{\frac}
\newcommand*{\avghaar}{\mathbb{E}_{_H}}

\newcommand*{\avg}{\mathbb{E}_{_U}}
\newcommand*{\avgdif}{\mathbb{E}_{_{U_i}}}

\newcommand*{\mc}{\mathcal}
\newcommand*{\dg}{\dagger}
\newcommand*{\mf}{\mathfrak}
\newcommand*{\mbb}{\mathds}
\newcommand*{\markov}{{\scriptscriptstyle{(\mathrm{M})}}}
\DeclareMathOperator{\tr}{tr}

\DeclareMathOperator{\swap}{\text{\scshape{swap}}}
\DeclareMathOperator{\FS}{\mf{S}}

\definecolor{quantpurple}{HTML}{53257F}
\definecolor{qorange}{HTML}{F75C03}
\definecolor{qgreen}{HTML}{04A777}
\definecolor{qred}{HTML}{D90368}
\definecolor{qmark}{HTML}{0E103D}

\begin{document}
\title[]{Almost Markovian processes from closed dynamics} 

\author{Pedro Figueroa--Romero}
\email[]{pedro.figueroaromero@monash.edu}
\author{Kavan Modi}
\author{Felix A. Pollock}

\affiliation{School of Physics and Astronomy, Monash University, Clayton, Victoria 3800, Australia}

\date{\today}

\begin{abstract}
It is common, when dealing with quantum processes involving a subsystem of a much larger composite closed system, to treat them as effectively memory-less (Markovian). While open systems theory tells us that non-Markovian processes should be the norm, the ubiquity of Markovian processes is undeniable. Here, without resorting to the Born-Markov assumption of weak coupling or making any approximations, we formally prove that processes are close to Markovian ones, when the subsystem is sufficiently small compared to the remainder of the composite, with a probability that tends to unity exponentially in the size of the latter. We also show that, for a fixed global system size, it may not be possible to neglect non-Markovian effects when the process is allowed to continue for long enough. However, detecting non-Markovianity for such processes would usually require non-trivial entangling resources. Our results have foundational importance, as they give birth to \textit{almost} Markovian processes from composite closed dynamics, and to obtain them we introduce a new notion of equilibration that is far stronger than the conventional one and show that this stronger equilibration is attained.
\end{abstract}


\maketitle

\section{Introduction}

The quest towards understanding how thermodynamics emerges from quantum theory has seen a great deal of recent progress~\cite{Gogolin, Goold}. Most prominently, the conundrum of how to recover irreversible phenomena that obey the second law of thermodynamics, starting from reversible and recurrent Schr\"odinger dynamics, has been rectified by considering an analogous dynamical equilibrium. This is known generically as \emph{equilibration on average}; it implies that time-dependent quantum properties evolve towards a certain fixed equilibrium value, and stay close to it for most times~\footnote{ Even stronger (but usually the one of interest) is the notion of equilibration in a finite time interval. The corresponding statement for a physical system replaces the term equilibration with \emph{thermalisation} and the equilibrium state with a \emph{thermal state}. For detail see e.g.~\cite{Gogolin}}.

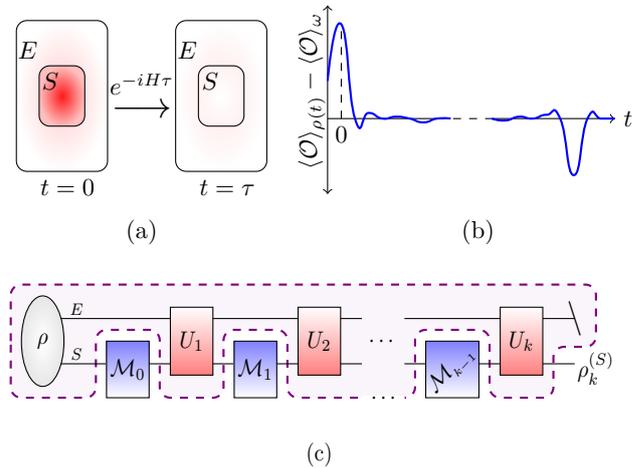
\begin{figure}[t]
      \begin{tikzpicture}[yscale=0.2, xscale=0.15, every node/.style={scale=0.925}]
      \shade[outer color=white, inner color=red!20!white, draw=black] (0,0) [rounded corners] rectangle (8,10);
      \shade[inner color=red!90!white, outer color=red!25!white, draw=black] (2,3) [rounded corners] rectangle (6,7);
      \node[scale=1.5] at (11,4) {$\longrightarrow$};
      \node[above] at (11,4.5) {$e^{-iH\tau}$};
      \shade[xshift=5.5in,outer color=white, inner color=red!10!white, draw=black] (0,0) [rounded corners] rectangle (8,10);
      \shade[xshift=5.5in,inner color=white, outer color=red!5!white, draw=black] (2,3) [rounded corners] rectangle (6,7);
      \node[below] at (4.5,0) {$t=0$};
      \node[below] at (18.5,0) {$t=\tau$};
      \node at (1,8) {$E$};
      \node at (3,6) {$S$};
      \node at (15,8) {$E$};
      \node at (17,6) {$S$};
      \node at (11,-4) {(a)};
      \end{tikzpicture}
~
    \begin{tikzpicture}[yscale=0.25, xscale=0.18, every node/.style={scale=0.925}]
    \draw[<->] (0,-4) -- (0,6);
    \draw[-] (0,0) -- (8,0);
    \draw[-, dashed] (8,0) -- (12,0);
    \draw[->] (12,0) -- (21,0);
    \draw [thick, blue] plot [smooth, tension=1] coordinates { (0,2) (1,5) (2,0) (3,0.3) (4,0) (5,0.1) (6,0) (7,-0.2) (8,0) (9,0)};
    \draw [thick, blue] plot [smooth, tension=1] coordinates { (12,0) (13,-0.1) (14,0.1) (15,0) (16,0.2) (17,0) (18,-3) (19,0) (20,0) (21,0)};
    \node[label={[label distance=0.5cm,text depth=-1ex,rotate=90]right:$\langle\mc{O}\rangle_{\rho(t)}-\langle\mc{O}\rangle_{\omega}$}] at (-2.5,-5) {};
    \node[right] at (21,0) {$t$};
    \node[below] at (1,0) {$0$};
    \draw[-, dashed] (1,0) -- (1,5);
    \node at (11,-6) {(b)};
    \end{tikzpicture}

\vspace{0.15 in}

\begin{tikzpicture}[xscale=0.28,yscale=0.3, every node/.style={xscale=0.8,yscale=0.9}]
    \draw[-, dashed, thick, red!50!blue, rounded corners, fill=red!50!blue!4!white] (-2.5,0) -- (-2.5,2.5) -- (25,2.5) -- (25,-0.25) -- (23,-0.25) -- (23,-2.5) -- (20,-2.5) -- (20,0.5) -- (16.5,0.5) -- (16.5,-2.5) -- (10.5,-2.5) -- (10.5,0.5) -- (7.5,0.5) -- (7.5,-2.5) -- (4.5,-2.5) -- (4.5,0.5) -- (1.5,0.5) -- (1.5,-2.5) -- (-2.5,-2.5) -- (-2.5,0);
    \draw[-, line width=0.5mm, white] (16.01,-2.5) -- (14,-2.5);
    \node at (15.1,-2.55) {$\cdots$};
    \shade[inner color=white, outer color=black!10!white, draw=black] (-1,0) ellipse (1cm and 2cm);
    \node at (-1,0) {$\rho$};
    \draw[-] (-0.15,1) -- (5,1);
    \draw[-] (-0.15,-1) -- (2,-1);
    \node[above] at (0.6,0.8) {\scriptsize${E}$};
    \node[above] at (0.6,-1.2) {\scriptsize${S}$};
    \shade[top color=blue!50!white, bottom color=white, draw=black] (2,-2.5) rectangle (4,0);
    \node at (3,-1.25) {$\mc{M}_{0}$};
    \draw[-] (4,-1) -- (5,-1);
    \shade[bottom color=red!50!white, top color=white, draw=black] (5,-1.5) rectangle (7,1.5);
    \node at (6,0) {$U_{1}$};
    \draw[-] (7,1) -- (11,1);
    \draw[-] (7,-1) -- (8,-1);
    \shade[top color=blue!50!white, bottom color=white, draw=black] (8,-2.5) rectangle (10,0);
    \node at (9,-1.25) {$\mc{M}_{1}$};
    \draw[-] (10,-1) -- (11,-1);
    \shade[bottom color=red!50!white, top color=white, draw=black] (11,-1.5) rectangle (13,1.5);
    \node at (12,0) {$U_2$};
    \node at (15,0) {$\cdots$};
    \draw[-] (13,1) -- (14,1);
    \draw[-] (13,-1) -- (14,-1);
    \draw[-] (16,1) -- (20.5,1);
    \draw[-] (16,-1) -- (17,-1);
    \shade[top color=blue!50!white, bottom color=white, draw=black] (17,-2.5) rectangle (19.5,0);
    \node[rotate=45] at (18.25,-1.25) {$\mc{M}_{_{k-1}}$};
    \draw[-] (19.5,-1) -- (20.5,-1);
    \shade[bottom color=red!50!white, top color=white, draw=black] (20.5,-1.5) rectangle (22.5,1.5);
    \node at (21.5,0) {$U_k$};
    \draw[-] (22.5,1) -- (24,1);
    \node at (24,1) {\Large\textbackslash};
    \draw[-] (22.5,-1) -- (24,-1);
    \node at (25,-1.2) {$\rho^{(S)}_k$};
    \node at (12,-5) {(c)};
    \end{tikzpicture}
\caption{\textbf{Weak and strong equilibration.} (a) A notion of subsystem equilibration: $S$ is in a nonequilibrium state at time $0$ and evolves jointly with $E$ according to Schr\"odinger's equation until it reaches an equilibrium state at some time $\tau$; more precisely, in (b) the expectation $\langle\,\cdot\,\rangle$ of an observable $\mc{O}$ on the state $\rho(t)$ differs only slightly from the corresponding one on the time-averaged state $\omega$ for most times, even if it eventually deviates greatly from it. In (c) we illustrate the generalization we consider in this work, where an initial joint state $\rho$ evolves unitarily between the application of measurements $\mc{M}_i$ on the system $S$, rendering the final reduced state $\rho_k^{(S)}$; the dotted box encloses the process tensor which describes the process independently of the (experimentally) controllable operations.}\label{Fig interventions}
\end{figure}

Equilibration on average has been widely studied for small non-degenerate subsystems of larger closed systems, which are usually taken to be (quasi) isolated, and for particular models (see e.g.~\cite{GarciaPintos}). In a typical setup, the closed dynamics of a system and environment, as illustrated in Fig.~\ref{Fig interventions}(a), is considered. Remarkably, it can be shown that the expectation value for \textit{any} observable $\mathcal{O}$ on the system alone lies in the neighbourhood of its time-averaged value for almost all times [Fig.~\ref{Fig interventions}(b)]. Yet, such equilibration results, in full generality, remain unsolved~\cite{Gogolin}; in particular, it is unknown whether they extend to correlations between observables at different times.

To see this, consider the case depicted in Fig.~\ref{Fig interventions}(c), where the subsystem is interrogated (by measuring, or otherwise manipulating it) multiple times as it evolves with the remainder of the closed system, which acts as an environment. In such a scenario, the flow of information between interventions must be taken into account; in particular, there may be intrinsic temporal correlations mediated by the environment, in other words \emph{memory} built up within the dynamics itself, independently of the external interventions, which are the hallmark of so-called \emph{non-Markovian} dynamics. Equilibration on average says that the process erases the information contained in the initial state of the system; however, there may still be non-Markovian memory of the initial state encoded in the temporal correlations between observables. It is therefore an open question whether this information is truly scrambled by the process and therefore inaccessible to the system at long times. Such memoryless dynamics are anticipated for the subparts of closed many-body dynamics~\cite{Srednicki1999}. This scrambling of information in correlations would represent a stronger notion of equilibration than the usual one, which only considers statistics of individual observables at a given time. Moreover, the absence of memory effects would generically imply the usual kind of equilibration.

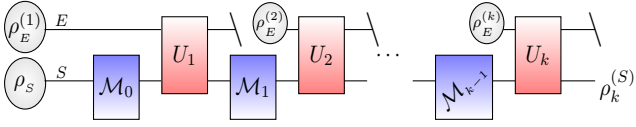
\begin{figure}[t]
\begin{tikzpicture}[xscale=0.3,yscale=0.34, every node/.style={xscale=0.8,yscale=0.9}]
    \draw[-, line width=0.5mm, white] (16,-3) -- (14,-3);
    \shade[inner color=white, outer color=black!10!white, draw=black] (-1,1.1) ellipse (0.9cm and 1cm);
    \shade[inner color=white, outer color=black!10!white, draw=black] (-1,-1.1) ellipse (0.9cm and 1cm);
    \node at (-1,1.1) {$\rho_{_E}^{(1)}$};
    \node at (-1,-1.1) {$\rho_{_S}$};
    \draw[-] (-0.15,1) -- (5,1);
    \draw[-] (-0.15,-1) -- (2,-1);
    \node[above] at (0.6,0.8) {\scriptsize${E}$};
    \node[above] at (0.6,-1.2) {\scriptsize${S}$};
    \shade[top color=blue!50!white, bottom color=white, draw=black] (2,-2.5) rectangle (4,0);
    \node at (3,-1.25) {$\mc{M}_{0}$};
    \draw[-] (4,-1) -- (5,-1);
    \shade[bottom color=red!50!white, top color=white, draw=black] (5,-1.5) rectangle (7,1.5);
    \node at (6,0) {$U_{1}$};
    \draw[-, rounded corners] (7,1) -- (8.25,1);
    \node at (8.25,1) {\Large\textbackslash};
    \draw[-, rounded corners] (9.75,1) -- (11,1);
    \shade[inner color=white, outer color=black!10!white, draw=black] (9.75,1.25) ellipse (0.75cm and 0.8cm);
    \node at (9.75,1.25) {$\scriptstyle{\rho_{_E}^{(2)}}$};
    \draw[-] (7,-1) -- (8,-1);
    \shade[top color=blue!50!white, bottom color=white, draw=black] (8,-2.5) rectangle (10,0);
    \node at (9,-1.25) {$\mc{M}_{1}$};
    \draw[-] (10,-1) -- (11,-1);
    \shade[bottom color=red!50!white, top color=white, draw=black] (11,-1.5) rectangle (13,1.5);
    \node at (12,0) {$U_2$};
    \node at (15,0) {$\cdots$};
    \draw[-, rounded corners] (13,1) -- (14.25,1);
    \node at (14.25,1) {\Large\textbackslash};
    \draw[-] (13,-1) -- (14,-1);
    \draw[-] (19.25,1) -- (20.5,1);
    \shade[inner color=white, outer color=black!10!white, draw=black] (19.25,1.25) ellipse (0.75cm and 0.8cm);
    \node at (19.275,1.25) {$\scriptstyle{\rho_{_E}^{(k)}}$};
    \draw[-] (16,-1) -- (17,-1);
    \shade[top color=blue!50!white, bottom color=white, draw=black] (17,-2.5) rectangle (19.5,0);
    \node[rotate=45] at (18.25,-1.25) {$\mc{M}_{_{k-1}}$};
    \draw[-] (19.5,-1) -- (20.5,-1);
    \shade[bottom color=red!50!white, top color=white, draw=black] (20.5,-1.5) rectangle (22.5,1.5);
    \node at (21.5,0) {$U_k$};
    \draw[-] (22.5,1) -- (24,1);
    \node at (24,1) {\Large\textbackslash};
    \draw[-] (22.5,-1) -- (24,-1);
    \node at (25,-1.2) {$\rho^{(S)}_k$};
    \end{tikzpicture}
    \caption{\textbf{Markovian dynamics.} A quantum circuit corresponding to the Markov approximation, which is mathematically equivalent to starting off with an uncorrelated $\rho_{_E}^{(1)}\otimes\rho_{_S}$ state and then artificially refreshing the environment (discarding and replacing with a new state) at each step (cf. Fig. \ref{Fig interventions}c)}\label{Markov approx circuit}
\end{figure}

Searching for this stronger notion of equilibration is not physically unmotivated. While, strictly speaking, non-Markovian dynamics is the norm for subsystems of a closed system evolving with a time-independent Hamiltonian~\cite{processtensor2, PhysRevA.92.022102}, Markov approximations are commonly made throughout physics (and elsewhere) to good effect. However, the standard approach to making this approximation in open quantum systems theory involves a series of assumptions about weak interactions with an environment sufficiently large and ergodic that the memory of past interactions gets practically lost~\cite{Schlosshauer2007, VegaAlonso}; this is mathematically equivalent (through an analogue of Stinespring's dilation theorem) to continually refreshing (discarding and replacing) the environment's state~\cite{CollisionModels}, i.e., artificially throwing away information from the environment (see Fig.~\ref{Markov approx circuit}). In reality, this is not usually how a closed system evolves, leading to the question: \textit{In the general case, how does dissipative Markovian dynamics emerge from closed Schr\"odinger evolution?} That this puzzle is similar to the ones concerning equilibration and thermalisation should not be too surprising, as thermalisation is often achieved by assuming a Markov process~\cite{spohn}.

Here, we give an initial answer to this deep foundational question. We show that, for the class of evolutions given by the Haar measure (typically involving an interaction of the subsystem with a significant portion of the remainder), above a critical time scale the dynamics of a subsystem is exponentially close to a Markov process as a function of the relative size of the subsystem compared with the whole. Moreover, our result does not assume weak coupling between the system and the environment, and we do not employ any approximations. Specifically, we use averages over the unitary group together with the concentration of measure phenomenon~\cite{Ledoux, milman, boucheron2013concentration, PopescuWinter, Li} to determine when a subsystem dynamics might typically be close to Markovian. Our approach is broadly similar to that of Refs.~\cite{PopescuWinter, Masanes} both in spirit and in terms of the mathematical tools employed. In other words, our main result gives a mechanism for equilibration starting from closed dynamics without employing any approximations.

Our results are possible thanks to recent efforts towards the understanding of general quantum processes, culminating in the development of the \emph{process tensor} framework~\cite{PhysRevLett.101.060401, PhysRevA.80.022339, Costa}, which allows for the complete and compact description of an open system's evolution under the influence of a series of external manipulations or measurements~\cite{Modi, processtensor}.

Crucially, the process tensor framework leads to an unambiguous distinction between Markovian and non-Markovian dynamics, analogous to the usual one for classical stochastic processes. In this sense, it subsumes other approaches to quantum non-Markovianity (when compared with respect to the same process), including those based on increasing trace-distance distinguishability~\cite{Breuer2015}. It allows for a precise quantification of memory effects (including across multiple time points)~\cite{processtensor2}, that we employ here as a natural setting to explore temporal correlations and the role that they play in a generalized, or strong, notion of equilibration.

\section{Quantum processes with memory: the process tensor}
We begin with the motivation of witnessing equilibration effects through a sequence of measurements on the system. We consider a closed environment-system ($ES$) with finite dimensional Hilbert space $\mc{H}_{ES}=\mathbb{C}^{d_E}\otimes\mathbb{C}^{d_S}$ initialised in a pure joint state $\rho$. The system $S$ is acted on with an initial measurement operation $\mc{M}_0$ followed by a series of subsequent measurement operations $\{\mc{M}_i\}_{i=1}^{k-1}$ at a sequence of fixed times, between which it evolves unitarily with the environment $E$; these operations are represented by trace non-increasing completely positive (CP) maps, and they include information about how the system is affected by the measurement procedure. Our results also hold for more general operations than just measurements, including unitary transformations or any other experimental manipulations of the system, but we focus on the former in order to be concrete. This scenario is depicted in Fig.~\ref{Fig interventions}(c). Formally, a $k$-step process tensor is a completely positive causal~\footnote{ Meaning that the output of a process tensor at time $k$ does not depend on deterministic inputs at later times $K>k$.} map $\mc{T}_{k:0}$ from the set $\{\mc{M}_i\}_{i=0}^{k-1}$ to output states, $\rho^\prime_{k}=\mc{T}_{k:0}[\{\mc{M}_i\}_{i=0}^{k-1}]$.

\begin{figure}[ht]
\centering
    \begin{tikzpicture}[thick, scale=0.3, every node/.style={scale=0.95}]
    \shade[inner color=white, outer color=black!10!white, draw=black] (-1,0) ellipse (1cm and 2cm);
    \node at (-1,0) {$\rho$};
    \draw[-] (-0.15,1) -- (2,1);
    \draw[-] (-0.15,-1) -- (2,-1);
    \node[above] at (1,0.8) {\scriptsize${E}$};
    \node[above] at (1,-1.2) {\scriptsize${S}$};
    \shade[bottom color=red!50!white, top color=white, draw=black] (2,-1.5) rectangle (4,1.5);
    \node at (3,0) {$U_{0}$};
    \draw[-] (4,1) -- (6,1);
    \draw[-] (4,-1) -- (6,-1);
    \shade[bottom color=red!50!white, top color=white, draw=black] (6,-1.5) rectangle (8,1.5);
    \node at (7,0) {$U_{1}$};
    \draw[-] (8,1) -- (10,1);
    \draw[-] (8,-1) -- (10,-1);
    \node at (11,0) {$\cdots$};
    \draw[-] (12,1) -- (14,1);
    \draw[-] (12,-1) -- (14,-1);
    \shade[bottom color=red!50!white, top color=white, draw=black] (14,-1.5) rectangle (16,1.5);
    \node at (15,0) {$U_{k}$};
    \draw[-] (16,1) -- (17,1);
    \node at (17,1) {\Large\textbackslash};
    \draw[-] (16,-1) -- (18,-1);
    \draw[-] (3,-3) -- (10,-3);
    \draw[-] (3,-5) -- (10,-5);
    \node at (11,-4) {$\cdots$};
    \draw[-] (12,-3) -- (18,-3);
    \draw[-] (12,-5) -- (18,-5);
    \draw (3,-3) arc (90:270:1);
    \node[left] at (2.25,-4) {$\Psi_{A_1B_1}$};
    \draw[-] (3,-7) -- (10,-7);
    \draw[-] (3,-9) -- (10,-9);
    \node at (11,-8) {$\cdots$};
    \draw[-] (12,-7) -- (18,-7);
    \draw[-] (12,-9) -- (18,-9);
    \draw (3,-7) arc (90:270:1);
    \node[left] at (2.25,-8) {$\Psi_{A_2B_2}$};
    \node[left] at (2.25,-10.5) {$\vdots$};
    \draw[-] (3,-12) -- (10,-12);
    \draw[-] (3,-14) -- (10,-14);
    \node at (11,-13) {$\cdots$};
    \node[left] at (11.5,-10.5) {$\vdots$};
    \draw[-] (12,-12) -- (18,-12);
    \draw[-] (12,-14) -- (18,-14);
    \draw (3,-12) arc (90:270:1);
    \node[left] at (2.25,-13) {$\Psi_{A_kB_k}$};
    \shade[outer color=red!50!blue!4!white, inner color=white, draw=black] (18,0) [rounded corners] rectangle (20.5,-15);
    \node[rotate=55] at (19.25,-7) {$\Upsilon_{k:0}$};
    \draw[-, blue!60!white] (5,-1) -- (5,-3);
    \node[blue!60!white] at (5,-1) {$\times$};
    \node[right] at (4.8,-2.1) {\scriptsize$\mc{S}_1$};
    \node[blue!60!white] at (5,-3) {$\times$};
    \draw[blue!60!white,-] (9,-1) -- (9,-7);
    \node[blue!60!white] at (9,-1) {$\times$};
    \node[right] at (8.8,-6.1) {\scriptsize$\mc{S}_2$};
    \node[blue!60!white] at (9,-7) {$\times$};
    \draw[blue!60!white,-] (13,-1) -- (13,-12);
    \node[blue!60!white] at (13,-1) {$\times$};
    \node[right] at (12.8,-11.1) {\scriptsize$\mc{S}_k$};
    \node[blue!60!white] at (13,-12) {$\times$};
    \node[above] at (3.25,-3.2) {\scriptsize$A_1$};
    \node[above] at (3.25,-5.2) {\scriptsize$B_1$};
    \node[above] at (3.25,-7.2) {\scriptsize$A_2$};
    \node[above] at (3.25,-9.2) {\scriptsize$B_2$};
    \node[above] at (3.25,-12.2) {\scriptsize$A_k$};
    \node[above] at (3.25,-14.2) {\scriptsize$B_k$};
    \end{tikzpicture}
\caption{\textbf{Choi state of a process tensor.} A quantum circuit diagram for the Choi state of a $k$-step process tensor under joint $ES$ unitary evolution: the final state $\Upsilon$ is a many-body quantum state capturing all the properties of the process.}\label{PT circuit diagram}
\end{figure}
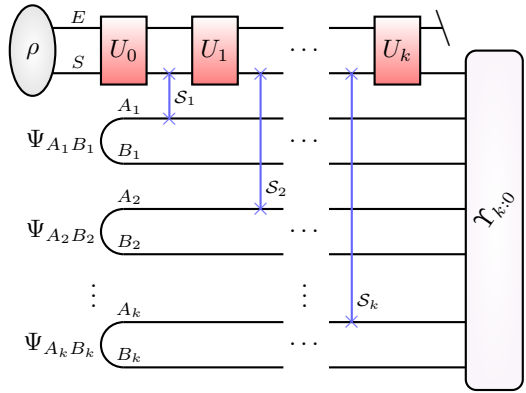

It must be stressed, however, that although experimental interventions play an important role in the reconstruction of a process through a generalized quantum process tomography, they do not determine its nature, i.e., the process tensor itself accounts precisely for the intrinsic dynamics occurring independently of the control of the experimenter. Moreover, like any other CP map, the process tensor can be expressed as a (many-body) quantum state through the Choi-Jamio\l{}kowski isomorphism~\cite{Choi1975, Jamiol1972,processtensor}. For a $k$-step process tensor, the corresponding (normalized) Choi state $\Upsilon_{k:0}$ can be obtained by swapping the system with one half of a maximally entangled state $\Psi_{A_iB_i}\equiv|\psi^+\rangle\!\langle\psi^+|\equiv\sum_{j,\ell=1}^{d_S}|jj\rangle\!\langle{\ell\ell}|/d_S$ at each time $i$~\cite{Simon2017}. The $2k$ ancillary systems $A_1,B_1,\cdots{A}_k,B_k$ have Hilbert space dimension $\dim(\mc{H}_A)=\dim(\mc{H}_B)=d_S$~\footnote{ The dimensions of each $A_i$, $B_i$ must be fixed to $d_S$ to make the swap operator well defined.} and the system is swapped with an $A$ subsystem at each time. With this convention, the Choi state of the process tensor can be seen to take the form
\begin{gather}
\begin{split}
&\Upsilon_{k:0}=\tr_E[\,\mc{U}_{k:0}(\rho\otimes\Psi^{\otimes{k}})\,\mc{U}^\dg_{k:0}],\\ \mbox{with} \quad
&\mc{U}_{k:0}\equiv(U_k\otimes\mbb1)\mc{S}_k\cdots(U_1\otimes\mbb1)\mc{S}_1(U_0\otimes\mbb1),
\end{split}
\label{choi evolution}
\end{gather}
where all identity operators act on the ancillary systems, $U_i$ is the $ES$ unitary operator that evolves the joint system following the control operation at time $i$ and the $\mc{S}_i$ are swap operators between system $S$ and the ancillary system $A_i$. The corresponding quantum circuit is depicted in Fig.~\ref{PT circuit diagram}.

Throughout the paper, we write $\Upsilon$ for the Choi state of a $k$-step process unless otherwise specified.

\section{An unambiguous measure of non-Markovianity}
Among other important properties, the process tensor leads to a well-defined Markov criterion, from which it is possible to construct a family of operationally meaningful measures of non-Markovianity, many of which can be stated simply as \emph{distances} between a process tensor's Choi state $\Upsilon$ and that of a corresponding Markovian one $\Upsilon^\markov$. The latter must take the form of a tensor product of quantum maps $\mc{E}_{i:i-1}$ connecting adjacent pairs of time steps: $\Upsilon^\markov = \bigotimes_{i=1}^k\mc{E}_{i:i-1}$~\cite{Simon2017}. 

In~\cite{processtensor2}, a such measure of non-Markovianity was first introduced through the relative entropy between a process and its corresponding Markovian one, $\mc{R}(\Upsilon\|\Upsilon^\markov)\equiv\tr[\Upsilon(\log\Upsilon-\log\Upsilon^\markov)]$, minimized over the latter. Here, however, in analogy with other studies on equilibration, we choose the measure of non-Markovianity defined in terms of the trace distance $\mc{D}$ as
\begin{gather}
\mc{N} \equiv\min_{\Upsilon^\markov}\mc{D}(\Upsilon,\Upsilon^\markov)
\equiv\f{1}{2}\min_{\Upsilon^\markov}\|\Upsilon-\Upsilon^\markov\|_1
\label{def nonMarkov trDistance}
\end{gather}
where $\|X\|_1\equiv\tr\sqrt{XX^\dg}$ is the trace norm (or Schatten 1-norm); in particular, this is related to relative entropy through the so-called quantum Pinsker inequality, $\mc{R}(\Upsilon\|\Upsilon^\markov)\geq2\,\mc{D}^2(\Upsilon,\Upsilon^\markov)$. The trace distance is a natural choice, as it represents an important distinguishability measure and has played a central role in the discovery of several major results on typicality~\cite{PopescuWinter} and equilibration~\cite{Gogolin, ShortFarrelly, Masanes} (or lack thereof, e.g.~\cite{PhysRevLett.106.040401}). It has an experimental importance for comparing quantum processes~\cite{PhysRevA.71.062310} and is naturally related to other important distinguishability measures such as the diamond norm or the fidelity. However, it is important to point out that the above measure is fundamentally different (and more powerful) than other measures of non-Markovianity that have been proposed in recent years, including one that employs trace distance~\cite{Breuer2015}. The measure we use here captures \textit{all} non-Markovian features across multiple time steps and provides an operational meaning for non-Markovianity. Equipped with this measure, we are in a position to determine what value it takes for a typical process, answering the question posed at the beginning of this manuscript.

\section{Typicality of Markovian processes}\label{sec: Main Typicality}
In order to quantify non-Markovianity, as defined in Eq.~\eqref{def nonMarkov trDistance}, for a typical process, we first need a concrete notion of typicality. There are many ways in which one could sample from the space of quantum processes, but as we are particularly interested in the typicality (or absence thereof) of Markovian processes independently of ad-hoc assumptions like weak coupling, we would like our measure to assign non-vanishing probabilities to mathematically generic unitary dynamics on the closed $ES$ system. 

Here, we achieve this by sampling the evolution from the unitarily invariant probability measure, the so-called Haar measure. This has the additional advantage of allowing us to average by means of random matrix theory techniques~\cite{ZhangMInt, Li, GuMoments, Collins, Puchala} and leads to the relatively straightforward application of concentration of measure results~\cite{Ledoux, milman, boucheron2013concentration}. Specifically, we use the Haar measure to sample two distinct types of $ES$ evolution:
\begin{enumerate}[leftmargin=*]
\item Random interaction: All $U_i$  independently chosen,
\item Constant interaction: $U_i=U_j,\quad \forall \leq{i}\neq{j}\leq{k}$,
\end{enumerate}
where $U_i$ represents the closed (unitary) $ES$ evolution between the application of measurement operation $\mc{M}_{i-1}$ and $\mc{M}_{i}$, as depicted in Fig.~\ref{Fig interventions}(c); the entire set enters into the process tensor through Eq.~\eqref{choi evolution}. In the first case, the global system will quickly explore its entire (pure) state space for any initial state. The second case corresponds more closely to what one might expect for a truly closed system, where the Hamiltonian remains the same throughout the process. These correspond to two extremes; more generally, the dynamics from step to step may be related but not identical.

We highlight that under this sampling method, and thus this notion of typicality, the interaction or information flow between all parts of the whole system plus environment is relevant, i.e. no parts of the environment dimension are superfluous. We will now show that this kind of dynamics is almost always close to Markovian.

\subsection{Main result}
Our main result states that, for a randomly sampled $k$-step quantum process $\Upsilon$ undergone by a $d_S$-dimensional subsystem of a larger $d_Ed_S$-dimensional composite, the probability for the non-Markovianity $\mc{N}$ to exceed a function of $k$, $d_S$ and $d_E$, that becomes very small in the large $d_E$ limit, itself becomes small in that limit. Precisely, we prove that, for an arbitrary $\epsilon>0$,
\begin{gather}
    \mathbb{P}[\mc{N}\geq\mc{B}_k(d_E,d_S)+\epsilon]\leq\mathrm{e}^{-
\mc{C}(d_E,d_S)\epsilon^2},\label{eq: main result}
\end{gather}
where $\mc{C}(d_E,d_S)=c \, d_Ed_S\left(\f{d_S-1}{d_S^{k+1}-1}\right)^2$ with $c=1/4$ for a constant interaction process and $c=(k+1)/4$ for a random interaction process. The function $\mc{B}_k(d_E,d_S)$ is an upper bound on the expected non-Markovianity $\mbb{E}[\mc{N}]$ whose details depend on the way in which processes are sampled.

\begin{figure}[t]
\centering
\begin{tikzpicture}[xscale=1.8, yscale=0.8, every node/.style={scale=1}]
    \shade[thick, scale=0.65, inner color=gray!5!white, outer color=white, draw=black]  (-2.225,0.4) ellipse (2.75cm and 2.5cm);
    \draw[-, qmark, very thick, dotted, decorate,decoration={snake, amplitude=.4mm, segment length=0.25in}, rotate around={150:(-0.75,0)}] (-0.7725,0) -- (0.7725,0);
    \node[qmark, right, above] at (-1,0.15) {$\Upsilon^\markov$};
    \node at (-1.425,-1.025) {$\Upsilon$};
    \draw[qorange,|<->|] (-1.32,-0.875) -- (-1.26,-0.65);
    \node[right, below, qorange] at (-1.2,-0.625) {$\epsilon$};
    \node[green!50!black] at (-2.33,0) {$\mc{B}_k$};
    \node[red!80!black] at (-1.9,0.2) {$\mathbb{E}[\mc{N}]$};
    \draw[-, red!80!black, very thick, dotted, rotate around={150:(-0.75,0)}] plot[smooth cycle, tension=.7] coordinates {(-1,0) (-0.8,0.3) (-0.6,0.4) (-0.3,0.6) (0,0.5) (0.4,0.8) (0.7,0.3) (1,0) (0.7,-0.3) (0.4,-0.6) (0,-0.5) (-0.3,-0.6) (-0.6,-0.4) (-0.8,-0.3)};
    \draw[-, green!50!black, very thick, dotted, rotate around={150:(-0.75,0)},scale=1.4] plot[smooth cycle, tension=.7] coordinates {(-1,0) (-0.8,0.3) (-0.6,0.4) (-0.3,0.6) (0,0.5) (0.4,0.8) (0.7,0.3) (1,0) (0.7,-0.3) (0.4,-0.6) (0,-0.5) (-0.3,-0.6) (-0.6,-0.4) (-0.8,-0.3)};
    \draw[blue!80!black, semithick, |<->|] (-1.415,-0.83) -- (-1.15,0.16);
    \node[right] at (-1.5, -0.04) {$\color{blue!80!black}{\mc{N}}$};
    \node[scale=1] at (-1.5,-2) {(a)};
    \end{tikzpicture}
    ~
    \begin{tikzpicture}[xscale=1.8, yscale=0.8, every node/.style={scale=1}]
    \shade[thick, scale=0.65, inner color=white, outer color=white, draw=black]  (-2.225,0.4) ellipse (2.75cm and 2.5cm);
    \foreach\i in {0,0.02,...,0.7} {
        \fill[opacity=\i*0.1, blue, rotate around={60:(-1,-1)}] (0,0) ellipse ({0.7-\i} and {1.8-1.8*\i});
      }
    \draw[qmark, -, very thick, dotted, decorate,decoration={snake, amplitude=.4mm, segment length=0.25in}, rotate around={150:(-0.75,0)}] (-0.7725,0) -- (0.7725,0);
    \node[qmark, right, above] at (-0.9,0.35) {$\Upsilon^\markov$};
    \node at (-1.5,-1) {$\Upsilon$};
    \node at (-1.5,-2) {(b)};
    \end{tikzpicture}
    \caption{\textbf{Concentration of Markovian processes in large environments.} (a) A geometric \emph{cartoon} of our main result in a space of process tensors: the probability of the non-Markovianity $\mc{N}$ of deviating from $\mc{B}_k$ by some $\epsilon>0$ decreases exponentially in $\epsilon^2$. In (b), quantum processes $\Upsilon$ on large dimensional environments (such that $d_E\gg{d}_S^{2k+1}$) concentrate around the Markovian ones $\Upsilon^\markov$.}\label{main cartoon}
\end{figure}
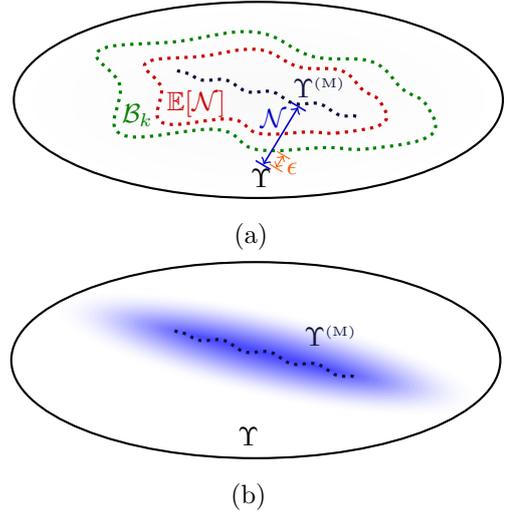

The proof is presented in full in Appendix~\ref{Appendix: Main Result}, and it can be outlined as follows: we first bound the difference between $\mc{N}$ for two different processes in terms of the distance between the unitaries used to generate them; Levy's lemma~\cite{Ledoux,boucheron2013concentration,PopescuWinter} then states that the fraction of processes with non-Markovianity more than $\epsilon$ away from the expectation value (which we upper-bound by $\mathcal{B}_k$) is bounded by a concentration function. By considering the geometry of the spaces of unitaries we are sampling from, we are able to use the exponential function appearing on the right hand side of Eq.~\eqref{eq: main result}. This implies that our result holds particularly for this class of evolutions where all parts of system and environment effectively interact, in contrast with some commonly considered open systems models.

Our result is meaningful when both $\mc{B}_k+\epsilon$ and $\mathrm{e}^{-\mc{C}\epsilon^2}$ are small; the latter is fulfilled in the \emph{small subsystem} or \emph{large environment} limit, which in our setting means $d_E\gg{d}_S^{2k+1}$. We may also state the minimal value for $\epsilon$ that, assuming the large environment limit, renders both sides small, i.e., such that $\epsilon^2d_E\gg1\gg\epsilon$; this is fulfilled for $\epsilon=d_E^{-1/3}$~\footnote{Similar to~\cite{PopescuWinter}, here we look for an $x>0$ such that $\epsilon=d_E^{-x}$ and $\epsilon^2d_E=d_E^x$}. A geometrical cartoon to illustrate the result is presented in Fig.~\ref{main cartoon}.

This is the case because the upper bound $\mc{B}_k \geq \mathbb{E}(\mc{N})$ is given by
\begin{gather}
\mc{B}_k(d_E,d_S)\equiv\begin{cases}
\f{\sqrt{d_E\,\mathbb{E}[\tr(\Upsilon^2)]-x}+y}{2}
&\text{if}\quad d_E<d_S^{2k+1}
\\[0.1in]
\f{\sqrt{d_S^{2k+1}\mathbb{E}[\tr(\Upsilon^2)]-1}}{2}&\text{if}
\quad d_E\geq{d}_S^{2k+1},\end{cases}\label{avg upper bound}\end{gather}
with $x\equiv\f{d_E}{d_S^{2k+1}}\left(1+y\right)$, $y \equiv 1-\f{d_E} {d_S^{2k+1}}$. The size of $\mc{B}_k$ depends entirely on the size of the average purity of the process tensor $\mathbb{E}[\tr(\Upsilon^2)]$, which we have computed analytically for the case of Haar-randomly sampled evolutions (which equivalently could be sampled from $(2k+2)$-unitary designs~\cite{chaosYoshida, Brandao2016}, as we will discuss below) in Appendix~\ref{sec:Purity Averages}.

The purity $\tr(\Upsilon^2)$ is a quantifier of the mixedness (uniformity of eigenvalues) of a positive operator. When computed on reduced states of bipartite systems it can also serve as a quantifier of entanglement. In Appendix~\ref{appendix: avg processes}, we analytically compute the expected purity of the Choi state of a process tensor $\mathbb{E}[\Upsilon^2]$ in both the constant and the random interaction pictures, which can be directly translated as a quantifier for \emph{noisiness} of the quantum process itself and entanglement between system and environment. The expressions take the form
\begin{gather}
    \avgdif[\tr(\Upsilon^{2})]=\f{d_E^2-1}{d_E(d_Ed_S+1)}\left(\f{d_E^2-1}{d_E^2d_S^2-1}\right)^k+\f{1}{d_E},\label{average purity ergodic}
\end{gather}
for the random interaction picture and
\begin{gather}
    \avg[\tr(\Upsilon^2)]=d_S^{-2k}\hspace*{-0.1in}\sum_{\sigma,\tau\in\mc{S}_{2k+2}}\hspace*{-0.1in}\mathrm{Wg}(\tau\sigma^{-1})\,\varrho\,_\tau\,\Delta_{k,\sigma,\tau}^{(d_E,d_S)},\label{average time independent purity}
\end{gather}
in the constant interaction case, where $\mc{S}_n$ is the symmetric group over $n$ elements, $\mathrm{Wg}$ is known as the Weingarten function~\cite{GuMoments}, $\varrho_\tau$ is a product of entries of $\rho$ depending on permutations $\tau$ and $\Delta$ is a product, scaling with $k$, of monomials in $d_E$ and $d_S$ depending on permutations $\sigma$ and $\tau$.

\subsection{Limiting cases}
In both the constant and random interaction cases, $\mc{B}_k$ is a well-behaved rational function of $d_E$, $d_S$ and $k$. Eq.~\eqref{average time independent purity} takes a non-trivial form mainly because of the $\mathrm{Wg}$ function (which is intrinsic to the Haar-unitary averaging in the constant interaction picture). However, due to results found in~\cite{GuMoments,Collins}, we can still study analytically the behaviour of the bound $\mc{B}_k$ for both cases in the two following limits.

In the large environment limit,
\begin{gather}
    \lim_{d_E\to\infty}\mbb{E}[\tr(\Upsilon^{2})]=\f{1}{d_S^{2k+1}},
\end{gather}
which corresponds to the purity of the maximally mixed state. Note that the averaging occurs after computing the purity; that is, independently of which case is considered. Every process sampled will be indistinguishable from the maximally noisy (and hence Markovian) one, with probability that tends to one as $d_E$ is increased:
\begin{gather}\lim_{d_E\to\infty}\mbb{E}[\mc{N}]=\lim_{d_E\to\infty}\mc{B}_k(d_E,d_S)=0.\end{gather}
Furthermore, from Eq.~\eqref{average purity ergodic} it is seen that it does so at a rate $\mc{O}(1/d_E)$ in the random interaction case.

The other interesting limiting case is the one where the $ES$ dimension is fixed, but the number of time steps is taken to be very large. The resulting process encodes all high order correlation functions between observables over a long period of time. Again for both cases, the expected purity in this limit goes as
\begin{gather}
    \lim_{k\to\infty}\mbb{E}[\tr(\Upsilon^{2})]=\f{1}{d_E},
\end{gather}
which corresponds to the maximally mixed purity of the environment. This implies that the Choi state of the full $ES$ unitary process is maximally entangled between $S$ and $E$. In this limit, we get correspondingly
\begin{gather}\lim_{k\to\infty}\mbb{E}[\mc{N}]\leq\lim_{k\to\infty}\mc{B}_k(d_E,d_S)=1,\end{gather}
meaning that nothing can be said about typical non-Markovianity; a typical process in this limit could be highly non-Markovian. Indeed, we expect this to be the case, since the finite-dimensional $ES$ space will have a finite recurrence time.

Both our main result in Eq.~\eqref{eq: main result}, and the bound on non-Markovianity, Eq.~\eqref{avg upper bound}, generalize well-known results for quantum states, i.e., when $k=0$. In this case, our main result reduces to the usual one for the typicality of maximally mixed states (or bipartite maximally entangled states), with $\mc{C}(d_E,d_S)=d_Ed_S/4$ (see e.g.~\cite{AlmostAllPureMME}). As we also detail in Appendix~\ref{sec:Purity Averages}, when we take the Haar measure, in either case, we recover the average purity
\begin{equation}
    \mbb{E}[\tr(\rho_S^{2})]=\f{d_E+d_S}{d_Ed_S+1},
\end{equation}
where $\rho_S\equiv\Upsilon_{0:0}=\tr_E(U\rho\,U^\dg)$, as found in~\cite{Lubkin, PageEntropy, LloydPagels, ScottCaves, Li, Giraud, Pasquale}. This then leads to
\begin{gather}
    \mbb{E}\left[\mc{D}\left(\rho_S,\f{\mbb1}{d_S}\right)\right]\leq\f{1}{2}\sqrt{\f{d_S^2-1}{d_Ed_S+1}},
\end{gather}
which goes as $\sqrt{d_S/4d_E}$ when $d_E\gg{d}_S$, and is also a standard result \cite{AlmostAllPureMME}.

\subsection{Numerical examples}
\begin{figure}[t]
     \includegraphics[width=0.475\textwidth]{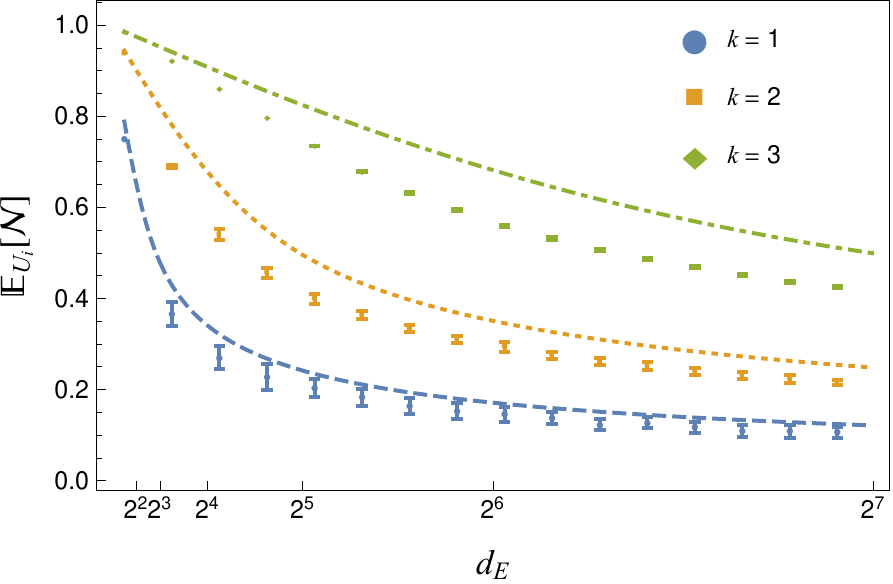}
     \caption{\textbf{Average non-Markovianity $\avgdif[\mc{N}]$ of a random interaction process for a qubit in the environment dimension $d_E$ at fixed time steps $k$}. Discrete values are shown for numerical averages over $\lfloor40/k\rfloor$ randomly generated process tensors $\Upsilon$ at time steps $k=1,2,3$ and with fixed $d_S=2$; error bars denote the standard deviation due to sampling error. The lines above each set of points denote the upper bound $\mc{B}_k(d_E,2)$. Process tensors were generated by sampling Haar random unitaries according to~\cite{Mezzadri}.}\label{numPlot}
\end{figure}

To support our results, we sampled Choi states $\Upsilon$ numerically in the random interaction case and computed their corresponding average non-Markovianity $\mathbb{E}[\mc{N}]$ as a function of environment dimension $d_E$ for a fixed system dimension $d_S=2$, obtaining the behaviour shown in Fig.~\ref{numPlot}. For constant interaction the numerical results are practically indistinguishable from those in the random case, but as mentioned, the analytical bound $\mc{B}_k$ is much harder to compute exactly. This suggests that either a simpler bound exists or that it might be possible to simplify the one we have obtained. As expected, our numerical results fall within the bound $\mc{B}_k(d_E,d_S=2)$ and they behave similarly; we notice that the bound in general seems to be somewhat loose, and become loosest when $d_E\simeq d_S^{2k+1}$, implying that non-Markovianity might be hard to detect even when not strictly in the large environment limit. However, it does saturate rapidly as $k$ increases.

So far, our results are valid for process tensors constructed with Haar random unitaries at $k$ evenly spaced steps; we are effectively considering a strong interaction between system and environment which rapidly scrambles quantum information in both~\cite{chaosYoshida}. The mechanism for equilibration is precisely that of dephasing~\cite{Gogolin}, or effectively, the scrambling of information on the initial state of the system. This suggests that, even when timescales will differ with the type of evolution considered, most physical evolutions fall within our result, with e.g. a weaker behavior in $k$.

Our results could potentially be extended to their analogues sampled from unitary $n$-designs, approximate or otherwise~\cite{Cotler2017}, that reproduce the Haar distribution up to the $n$-th moment. As mentioned above, our expression for the upper bound in Eq.~\eqref{avg upper bound} is identical for samples from a $2k+2$-designs, and, if the non-Markovianity $\mc{N}$ could be approximated by a polynomial function of $U$, Eq.~\eqref{eq: main result} would also hold for $n$-designs with large enough $n$~\cite{Low}. This type of distribution has recently been shown to be approximated by a wide class of physical evolutions~\cite{PhysRevX.7.021006}. In particular, the unitary circuits involved in complex quantum computations typically look Haar random~\cite{Brandao2016}; if it holds in this case, our result would therefore suggest that the behaviour of components of a quantum computer would not typically exhibit non-Markovian memory (at least due to the computation itself).  Furthermore, even when it's known that ensembles $\{e^{-iHt}\}_{t\geq0}$ for time-independent Hamiltonians $H$ cannot strictly become Haar random~\cite{chaosYoshida}, certain complex Hamiltonian systems can be treated as random up to a certain Haar-moment via unitary designs~\cite{PhysRevX.7.021006}.

These issues should be investigated further, however, we will now show that our results still hold at a coarse-grained level, where the intermediate dynamics corresponds to products of Haar random unitaries, which are not themselves Haar random.

\section{Observing non-Markovianity}
The choice of unitaries going into the process tensor in the previous sections (in our case drawn from the Haar measure), dictates a time scale for the system, up to a freely chosen energy scale. However, we can straightforwardly construct process tensors on a longer, coarse-grained time scale by simply allowing the system to evolve (with an identity operation) between some subset of time steps. In fact, process tensors at all time scales should be related in this way to an underlying process tensor with an infinite number of steps~\cite{milz_kolmogorov_2017}. 

To see that our main result directly applies to any process tensor that can be obtained through coarse graining, we again consider the definition of our non-Markovianity measure, given in Eq.~\eqref{def nonMarkov trDistance}. Consider the coarser grained process tensor $\Upsilon_{k:0/\{i\in[0,k-1]\}} \equiv \Upsilon_c$, where a subset of control operations $\{\mc{M}_i\}_{i\in[0,k-1]}$ are replaced by identity operations (i.e., the system is simply left to evolve). Letting $\mc{N}_\text{coarse} \equiv \min_{\Upsilon_c^\markov} \mc{D}(\Upsilon_c, \Upsilon_c^\markov)$, we have
\begin{gather}
\mc{N}_\text{coarse}\leq\mc{N},
\label{eq:coarsegraining}
\end{gather}
since the set of allowed $\Upsilon_c^\markov$ strictly contains the allowed $\Upsilon^\markov$ at the finer-grained level. This renders the new process less distinguishable from a Markovian one, i.e., coarse-graining can only make processes more Markovian.

The physical intuition behind this result is that the amount of information which can be encoded in a coarse grained process tensor is strictly less than that in its parent process tensor. In fact, this is a key feature of non-Markovian memory: the memory should decrease under \textit{coarse graining}. On the other hand, due to the same reasoning, we cannot say anything about finer-grained dynamics. One approach to tackling this issue would be to choose a different sampling procedure which explicitly takes scales into account. However, we leave this for future work.

There is another important limitation for observing non-Markovianity. The operational interpretation of the trace distance, discussed in the previous section, implies that observing non-Markovianity requires applying a measurement that is an eigen-projector operator of $\Upsilon-\Upsilon^\markov$. The optimal measurement will, in general, be entangled across all time steps. In practice, this is hard to achieve and typically one considers a sequence of local measurements $m\in\mathbb{M}$. In general, for an any set of measurements $\mathbb{M}$ we can define a restricted measure of non-Markovianity detectable with that set: $\mc{D}_{\mathbb{M}} (\Upsilon,\Upsilon^\markov)\equiv\max_{m\in\mathbb{M}} \f{1}{2} |\tr[m(\Upsilon-\Upsilon^\markov)]|  \leq \mc{D} (\Upsilon,\Upsilon^\markov)$, which means that the detectable non-Markovianity will be smaller. This is akin to the \textit{eigenstate thermalization hypothesis}~\cite{PhysRevLett.120.150603}, where all eigenstates of a physical Hamiltonian look uniformally distributed with respect to most `physically reasonable' observables. In our setting, this means that looking for non-Markovianity with observables that are local in time -- i.e., `physically reasonable' -- we  find almost no temporal correlations.

The locality constraint, along with monotonicity of non-Markovianity under coarse graining, have further important consequences for a broad class of open systems studies where master equations are employed~\cite{arXiv:1704.06204}. Since master equations usually only account for two-point correlations with local measurements, they will be insensitive to most of the temporal correlations being accounted for by our measure, leading to an even greater likelihood for their descriptions to be Markovian. We now discuss the broader implications of our results.

\section{Discussions and Conclusions}\label{sec: discussion}
Our results imply that it is fundamentally hard to observe non-Markovianity in a typical process and thus go some way to explaining the overwhelming success of Markovian theories.

Specifically, we have shown that, even when interacting strongly with the wider composite system, a subsystem will typically undergo highly Markovian dynamics when the rest of the system has a sufficiently large dimension, and that the probability to be significantly non-Markovian vanishes with the latter. Our main result formalizes the notion that in the large environment limit a quantum process, taken uniformly at random, will be almost Markovian with very high probability. This corroborates the common understanding of the Born-Markov approximation~\cite{Schlosshauer2007, VegaAlonso}, but, crucially, we make no assumptions about weak coupling between $E$ and $S$. Instead, in the Haar random interactions we consider, every part of the system typically interacts significantly with every part of $E$. This is in contrast to many open systems models, even those with superficially infinite dimensional baths, where the effective dimension of the environment is relatively small~\cite{tamascelli2018}; it can always be bounded by a function of time scales in the system-environment Hamiltonian~\cite{Luchnikov2018}, which could be encoded in a bath spectral density. Our result is also more general than the scenario usually considered, since it accounts for interventions and thus the flow of information between $S$ and $E$ across multiple times.

While it may still be possible to observe non-Markovian behaviour at a time scale that is smaller than the fundamental time scale set by the chosen unitaries, Eq.~\eqref{eq:coarsegraining} tells us that any coarse grained process will remain concentrated around the Markovian ones in the large environment limit. Otherwise, for larger and larger systems, one needs an ever increasing number of time steps, corresponding to higher order correlations, in order to increase the probability of witnessing non-Markovianity. However, even in this case, from the discussion in the previous section, we know that the measurement on this large number of times steps will be temporally entangled, which may also be difficult to achieve.

We have introduced a stronger notion of equilibration than the conventional one~\cite{Gogolin}, so that strong equilibration implies weak equilibration. Our main result shows that it is possible to attain strong equilibration, where the information of the initial state of the system does not exist in multi-time correlations, much less expectation values at different times. Our result can be interpreted as a mechanism for equilibration and thermalisation. That is, we have shown that the dynamics of the system alone is almost Markovian, and such processes have well defined fixed points~\cite{fixedpoint}. Even if we do not characterise the equilibrium state here, e.g. as done for quantum states under randomized local Hamiltonians~\cite{Cramer}, we may conclude that the state of the system approaches this fixed point and the expectation value for any observable will lie in the neighbourhood of the time-averaged value [as depicted in Fig.~\ref{Fig interventions}(b)] for most times. However, one point of departure between the two results is our reliance on Haar sampling.

More precisely, we have considered the case where the joint unitary evolution between each time step is drawn uniformly at random according to the Haar measure, for the two limiting cases of the random unitary operators at each step being the same or independent. This is, however, different from simply sampling the Choi state of the process Haar randomly. It is also not equivalent to picking Haar random states for the system at every time, as one would expect if the results of Ref.~\cite{PopescuWinter} were to hold independently at all times.

Nevertheless, tackling the issues we have pointed out, a notion of equilibration in general processes can potentially be made precise and other issues such as equilibration time scales can also be readily explored; the approach that we have presented constitutes a first significant step towards achieving this goal.

\begin{acknowledgments}
We are grateful to Simon Milz and Andrea Collevecchio for valuable discussions. PFR is supported by the Monash Graduate Scholarship (MGS) and the Monash International Postgraduate Research Scholarship (MIPRS). KM is supported through Australian Research Council Future Fellowship FT160100073.
\end{acknowledgments}

\onecolumn\newpage
\appendix

\section{The process tensor}\label{appendix: process tensor}
The Choi state of a process tensor with initial state $\rho$ is given by
\begin{gather}
    \Upsilon_{k:0}=\tr_E[\,\mc{U}_{k:0}(\rho\otimes\Psi^{\otimes{k}})\,\mc{U}^\dg_{k:0}],
\label{PT Choi state Def}
\end{gather}
with
\begin{gather}\mc{U}_{k:0}\equiv(U_k\otimes\mbb1)\mc{S}_k\cdots(U_1\otimes\mbb1)\mc{S}_1(U_0\otimes\mbb1),\end{gather}
where all identities are in the total ancillary system and the $U_i$ are $ES$ unitary operators at step $i$ (here the initial unitary $U_0$ is taken simply to randomize the initial state according to the Haar measure), and
\begin{gather}\mc{S}_i\equiv\sum_{\alpha,\beta}\mf{S}_{\alpha\beta}\otimes\mbb1_{_{A_1B_1\cdots{A}_{i-1}B_{i-1}}}\otimes|\beta\rangle\!\langle\alpha|\otimes\mbb1_{_{B_iA_{i+1}B_{i+1}\cdots{A}_kB_k}},\end{gather}
with $\mf{S}_{\alpha\beta}=\mbb1_E\otimes|\alpha\rangle\!\langle\beta|$. Notice that $\mf{S}_{\alpha\sigma}=\mf{S}_{\sigma\alpha}^\dg$, $\mf{S}_{ab}\mf{S}^\dg_{cd}=\delta_{bd}\mf{S}_{ac}$ and $\tr(\mf{S}_{ab})=d_E\delta_{ab}$. Also, $\Upsilon\in\mc{S}(\mc{H}_{S}\bigotimes_{i=1}^k\mc{H}_{A_iB_i})$, i.e. the resulting Choi process tensor state belongs to the $S$-ancillary system, which has dimension $d_S^{2k+1}$.

\section{Derivation of an upper-bound on non-Markovianity}\label{proof avg nonMark}
As a first approach, given the complexity of computing (averages over) the Markovian Choi state, we may upper bound this distance by the one with respect to the maximally mixed state (the noisiest Markovian process possible)
\begin{gather}\mc{N}\leq{D}(\Upsilon,d_S^{-(2k+1)}\mbb1).\label{TrDist 1st bound}\end{gather}
We may further bound this by considering the following cases separately.

\textbf{1. Case} $d_E<d_S^{2k+1}$: We notice that\textsuperscript{\footnote{One may see this by looking at the pure state ${|\Phi\rangle\!\langle\Phi|\equiv\mc{U}(\Theta\otimes\Psi^{\otimes{k}})\mc{U}^\dg}$ with Schmidt decomposition ${|\Phi\rangle=\sum_{i=1}^n\sqrt{\lambda_i}|e_is_i\rangle}$ where $n=\min(d_E,d_S^{2k+1})$.}} $\mathrm{rank}(\Upsilon)\leq{d_E}$. Letting $\gamma$ be the diagonal matrix of up to $d_E$ non-vanishing eigenvalues $\lambda_{\gamma_i}$ of the Choi state, we may write
\begin{gather}
\|\Upsilon-\f{\mbb1}{d_S^{2k+1}}\|_1=\sum_{i=1}^{d_E}\left|\lambda_{\gamma_i}-\f{1}{d_S^{2k+1}}\right|+\sum_{j=d_E+1}^{d_S^{2k+1}}\left|-\f{1}{d_S^{2k+1}}\right|=\|\gamma-\f{\mbb1_E}{d_S^{2k+1}}\|_1+1-\f{d_E}{d_S^{2k+1}},
\end{gather}
where $|\cdot|$ denotes the standard absolute value, so using the inequality ${\|X\|_1\leq\sqrt{\mathrm{dim}(X)}\|X\|_2}$ for a square matrix $X$ (which can be derived from the Cauchy-Schwarz inequality applied to the eigenvalues of $X$), where $\|X\|_2=\sqrt{\tr(XX^\dg)}$ is the Schatten 2-norm,
\begin{align}
\|\Upsilon-\f{\mbb1}{d_S^{2k+1}}\|_1 & \leq \sqrt{d_E}\|\gamma-\f{\mbb1_E}{d_S^{2k+1}}\|_2+1-\f{d_E}{d_S^{2k+1}} \nonumber \\
&=\sqrt{d_E\tr[\Upsilon^2]+\f{d^2_E}{d_S^{4k+2}}-\f{2 d_E} {d_S^{2k+1}}} +1-\f{d_E}{d_S^{2k+1}},
\end{align}
Furthermore, applying Jensen's inequality (in particular for the square-root, $\mathbb{E}[\sqrt{X}]\leq\sqrt{\mathbb{E}[X]}$) with expectation over evolution (either in the ergodic or time-independent case), this gives
\begin{gather}
\mathbb{E}\left[\mc{N}\right]\leq\f{1}{2}\left(\sqrt{d_E\mathbb{E}[\tr(\Upsilon^2)]+\f{d_E^2}{d_S^{4k+2}}-\f{2d_E}{d_S^{2k+1}}}+1-\f{d_E}{d_S^{2k+1}}\right).\label{up bound2}
\end{gather}

\textbf{2. Case} $d_E\geq{d}_S^{2k+1}$: This case is a small subsystem limit for most $k$. Directly applying ${\|X\|_1\leq\sqrt{\mathrm{dim}(X)}\|X\|_2}$ as before,
\begin{gather}
\|\Upsilon-\f{\mbb1}{d_S^{2k+1}}\|_1
\leq\sqrt{d_S^{2k+1}}\|\Upsilon-\f{\mbb1}{d_S^{2k+1}}\|_2 = \sqrt{d_S^{2k+1}\tr[\Upsilon^2] -1},
\end{gather}
Similarly, taking the average over evolution, by means of Jensen's inequality,
\begin{align}\mathbb{E}\left[\mc{N}\right]&\leq\f{1}{2}\sqrt{d_S^{2k+1}\mathbb{E}\left[\tr(\Upsilon^2)\right]-1}.\label{up bound1}\end{align}
This proves that the function
\begin{gather}
\mc{B}_k(d_E,d_S)\equiv\begin{cases}
\f{\sqrt{d_E\,\mathbb{E}[\tr(\Upsilon^2)]-x}+y}{2}
&\text{if}\quad d_E<d_S^{2k+1}
\\[0.1in]
\f{\sqrt{d_S^{2k+1}\mathbb{E}[\tr(\Upsilon^2)]-1}}{2}&\text{if}
\quad d_E\geq{d}_S^{2k+1},\end{cases}\end{gather}
with $x\equiv\f{d_E}{d_S^{2k+1}}\left(1+y\right)$, $y \equiv 1-\f{d_E} {d_S^{2k+1}}$, as given in the main text, provides an upper bound on the average non-Markovianity $\mathbb{E}[\mc{N}]$.

\section{Haar random averages}\label{appendix: Haar}
\subsection{The Haar measure}\label{Haar measure}
The Haar measure is the unique probability measure $\mu$ on any of the orthogonal, unitary or symplectic groups that is both left and right invariant, for our purposes meaning that $\int{f}(VU)\,d\mu(U)=\int{f}(UV)\,d\mu(U)=\int{f}(U)\,d\mu(U)$ for all unitaries $V$, with normalization $\int{d}\mu(U)=1$. A random pure state can thus be generated by taking $|\phi\rangle\in\mc{H}$ with $d=\dim(\mc{H})$, sampling a unitary matrix $U\in\mathrm{U}(d)$ according to the Haar measure and letting $|\phi\rangle\!\langle\phi|\equiv\phi\mapsto{U}\phi{U}^\dg$. Its average over the Haar measure is then given by
\begin{gather}\avghaar(\phi)\equiv\int\limits_{\mathrm{U}(d)}U\phi\,{U}^\dg\,d\mu(U),\end{gather}
which can also be seen as a so-called \emph{1-fold twirl} and be computed by the Schur-Weyl duality~\cite{ZhangMInt, chaosYoshida}. When we deal with averages over the Haar measure with several unitaries we will denote by $\avgdif$ and $\avg$ the ergodic and time-independent cases, respectively. 

\subsection{Schur-Weyl duality and the moments of the unitary group}\label{SW duality}
The Schur-Weyl duality allows to compute integrals of \emph{twirled} maps over the Haar measure given by
\begin{gather}\Phi^{(k)}(X)\mapsto\int\limits_{U(d)}{U}^{\otimes{k}}X({U}^{\otimes{k}})^\dg\,d\mu(U),\end{gather}
where $X\in\mc{H}^{\otimes{k}}$; in particular, the case $U^\dg{A}UX{U}^\dg{B}U$ can be taken to a twirled form and be evaluated as in e.g. in~\cite{ZhangMInt}, which we here use to compute the average purity of the Choi (ergodic) process tensor. In particular, $\Phi^{(k)}$ commutes with all $V^{\otimes{k}}$ such that $V\in\mathrm{U}(d)$, and by such property, the Schur-Weyl duality assures that also
\begin{align}\Phi^{(k)}(X)=\sum_{\sigma\in\mc{S}_k}f_\tau(X)\mc{P}_\tau,\end{align}
where the sum is over the symmetric group $\mc{S}_k$ on $k$ symbols, $\mc{P}_\tau$ is a permutation operator defined by $\mc{P}_\tau|x_1,\ldots,x_k\rangle=|x_{\tau(1)},\ldots,x_{\tau(k)}\rangle$ and $f_\tau$ is a linear function of $X$~\cite{chaosYoshida}. The $k=1$ case is well known~\cite{Li} and in general for any operator acting on $\mbb{C}^d$ results in a completely depolarizing channel,
\begin{gather}\Phi^{(1)}(X)=\mathbb{E}({X})=\f{\tr(X)}{d}\mbb{1}.\label{1-fold twirl}\end{gather}
A related result, which can equivalently be used to compute integrals over the unitary group for polynomials in the unitaries with the Haar measure, and which we also make use of in this work, is the one for the $k$-moments of the $d$-dimensional unitary group~\cite{GuMoments},
\begin{gather}\int\limits_{\mathrm{U}(d)}\prod_{\ell=1}^kU_{i_\ell{j}_\ell}U_{i^\prime_\ell{j}^\prime_\ell}^*\,d\mu(U)=\sum_{\sigma,\tau\in\mc{S}_k}\prod_{\ell=1}^k\delta_{i_\ell{i}^\prime_{\sigma(\ell)}}\delta_{j_\ell{j}^\prime_{\tau(\ell)}}\,\mathrm{Wg}(\tau\sigma^{-1},d),\label{k moments of U}\end{gather}
where $U_{ij}$ is the $ij$ component of $U\in\mathrm{U}(d)$ and $\mathrm{Wg}$ is known as the Weingarten function (here specifically for the unitary group)~\cite{Collins, Weingarten}. The Weingarten function can be defined in different ways, the most common being in terms of a sum over \emph{partitions} (or, equivalently, \emph{Young tableaux}) of $k$ and the so-called \emph{characters} of the symmetric group; it is a fairly complicated function to evaluate explicitly and we refer to~\cite{GuMoments, ZhangMInt} for the details, in any case it is simply a rational function in the dimension argument $d$ (particular cases are often given in the literature for small $k$, see e.g.~\cite{GuMoments, chaosYoshida}). An alternative is to perform numerical calculations, as a computer package in the Mathematica software was developed in~\cite{Puchala} that allows to numerically evaluate Eq.~\eqref{k moments of U} and the Weingarten function (a subsequent version for Maple is also now available in~\cite{IntHaar}).

Here we obtain the expected states $\mathbb{E}(\Upsilon)$ and expected purities $\mathbb{E}[\tr(\Upsilon^2)]$ over the Haar measure of Choi process tensor states $\Upsilon$ as defined by Eq.~\eqref{PT Choi state Def}. We compute these averages for both the ergodic $\avgdif$ and time-independent $\avg$ cases. The time-independent averages are written in terms of the $\mathrm{Wg}$ function; we express them explicitly for the simplest cases and we show that in the small subsystem limit they reduce to the corresponding ergodic averages.

The asymptotic behavior of Eq.~\eqref{k moments of U} boils down to that of the $\mathrm{Wg}$ function and in~\cite{GuMoments} it has been shown that
\begin{gather}\mathrm{Wg}(\sigma\in\mc{S}_n,d)\sim\f{1}{d^{2n-\#\sigma}},\,\,\text{as}\,\,d\to\infty,\label{Wg asympt result}\end{gather}
as a refinement of a result in~\cite{Collins}, where $\#\sigma$ is the number of cycles of the permutation $\sigma$ counting also \emph{fixed points} (assignments from an element to itself, $\sigma(x)=x$).

\section{Average Processes}\label{appendix: avg processes}
\subsection{Ergodic average process Choi state}\label{appendix: ergodic avg state}
The ergodic average Choi state of a $k$ step process is given by the maximally mixed state,
\begin{gather}\avgdif(\Upsilon_{k:0})=\f{1}{d_S^{2k+1}}\mbb1_{SA_1B_1\cdots{A}_kB_k}.\end{gather}
This can be seen by a simple guess, as there are $k+1$ independent integrals of a 1-fold twirl kind, and both the initial state and the $\FS$ operators have trace one. This is also an intuitive result, here corresponding to the \emph{noisiest} process possible.

The detail of the calculation is as follows.

From the definition of the Choi process tensor state~\eqref{PT Choi state Def}, in the ergodic case,
\begin{align}\Upsilon_{k:0}&=\tr_E\left[(U_k\otimes\mbb1)\mc{S}_k\cdots(U_1\otimes\mbb1)\mc{S}_1(U_0\otimes\mbb1)(\rho\otimes\Psi^{\otimes{k}})(U_0^\dg\otimes\mbb1)\mc{S}_1^\dg(U_1^\dg\otimes\mbb1)\cdots\mc{S}_k^\dg(U_k^\dg\otimes\mbb1)\right]\nonumber\\
&=\tr_E\sum\left[{U}_k\FS_{\alpha_k\beta_k}
\cdots{U}_1\FS_{\alpha_1\beta_1}U_0\rho
{U}_{0}^\dg\FS_{\gamma_1\delta_1}^\dg{U}_1^\dg\cdots\FS_{\gamma_k\delta_k}^\dg{U}_k^\dg\right]\nonumber\\
&\hspace{0.25in}\otimes[(|\beta_1\rangle\!\langle\alpha_1|\otimes\mbb1_{B_1})\Psi(|\gamma_1\rangle\!\langle\delta_1|\otimes\mbb1_{B_1})\otimes\cdots\otimes(|\beta_k\rangle\!\langle\alpha_k|\otimes\mbb1_{B_k})\Psi(|\gamma_k\rangle\!\langle\delta_k|\otimes\mbb1_{B_k})],\end{align}
where the sum is over repeated indices (Greek letters), hence by repeatedly evaluating the 1-fold twirl~\eqref{1-fold twirl}, with $d=d_Ed_S$ (also, the unitary groups are implicitly of dimension $d_Ed_S$), we get
\begin{align}&\avgdif(\Upsilon_{k:0})=\tr_E\sum\int\limits_{U_k\cdots{U}_0}U_k\FS_{\alpha_k\beta_k}
\cdots{U}_1\FS_{\alpha_1\beta_1}U_0\rho
{U}_{0}^\dg\FS_{\gamma_1\delta_1}^\dg{U}_1^\dg\cdots\FS_{\gamma_k\delta_k}^\dg{U}_k^\dg\nonumber\\
&\hspace{1.25in}\otimes[(|\beta_1\rangle\!\langle\alpha_1|\otimes\mbb1_{B_1})\Psi(|\gamma_1\rangle\!\langle\delta_1|\otimes\mbb1_{B_1})\otimes\cdots\otimes(|\beta_k\rangle\!\langle\alpha_k|\otimes\mbb1_{B_k})\Psi(|\gamma_k\rangle\!\langle\delta_k|\otimes\mbb1_{B_k})]\nonumber\\
&=\f{1}{d_S^kd}\tr_E\sum\int\limits_{U_{k-1}\cdots{U}_0}\langle{e}_{k}\beta_{k}|U_{k-1}\FS_{\alpha_{k-1}\beta_{k-1}}\cdots{U}_1\FS_{\alpha_1\beta_1}U_0\rho
{U}_{0}^\dg\FS_{\gamma_1\delta_1}^\dg{U}_1^\dg\cdots\FS_{\gamma_{k-1}\delta_{k-1}}^\dg{U}_{k-1}^\dg|e_k\delta_k\rangle\nonumber\\
&\hspace{1.25in}\mbb{1}_S\otimes|\beta_1\alpha_1\cdots\beta_k\alpha_k\rangle\!\langle\delta_1\gamma_1\cdots\delta_k\alpha_k|\nonumber\\
&=\f{d_E}{d_S^kd^2}\tr_E\sum\int\limits_{U_{k-2}\cdots{U}_0}\langle{e}_{k-1}\beta_{k-1}|U_{k-2}\FS_{\alpha_{k-2}\beta_{k-2}}\hspace{-0.1in}\cdots{U}_1\FS_{\alpha_1\beta_1}U_0\rho
{U}_{0}^\dg\FS_{\gamma_1\delta_1}^\dg{U}_1^\dg\cdots\FS_{\gamma_{k-2}\delta_{k-2}}^\dg{U}_{k-2}^\dg|e_{k-1}\delta_{k-1}\rangle\nonumber\\
&\hspace{1.25in}\mbb{1}_S\otimes|\beta_1\alpha_1\cdots\beta_{k-1}\alpha_{k-1}\rangle\!\langle\delta_1\gamma_1\cdots\delta_{k-1}\alpha_{k-1}|\otimes\mbb1_{A_kB_k}\nonumber\\
&\vdots\nonumber\\
&=\f{d_E^{k+1}}{d_S^kd^{k+1}}\mbb1_{SA_1B_1\cdots{A}_kB_k}\nonumber\\
&=\f{1}{d_S^{2k+1}}\mbb1_{SA_1B_1\cdots{A}_kB_k},\end{align}
where the Haar measure symbols are implied but omitted just to economize notation.

\subsection{Time-independent average process Choi state}\label{appendix: average state Ui=Uj}
The average Choi process tensor for $k$ steps in the time-independent case can be given in terms of a set $\{|s_i^{(\prime)}\rangle\}_{s_i=1}^{d_S}$ of $S$ system bases for $i=0,\ldots,k$ as
\begin{align}&\avg(\Upsilon_{k:0})=\f{1}{d_S^k}\sum_{\sigma,\tau\in\mc{S}_{k+1}}\rho_{_{\tau(0);0}}\mathrm{Wg}(\tau\sigma^{-1})\Delta_{k,\sigma,\tau}^{(d_E)}
|s_{\sigma(k)}\rangle\!\langle{s}_k|\bigotimes_{j=1}^k|s_{\sigma(j-1)}s^\prime_{\tau(j)}\rangle
\langle{s}_{j-1}s^\prime_{j}|,\label{average state Ui=Uj}\end{align}
with implicit sum over all repeated basis ($s^{(\prime)}_i$) indices, where here $\mc{S}_{k+1}$ is the symmetric group on $\{0,\ldots,k\}$, and with the definitions
\begin{align}\rho_{_{\tau(0);0}}&=\langle{e}^\prime_{\tau(0)}s^\prime_{\tau(0)}|\rho|e^\prime_0s^\prime_0\rangle,\label{phi notation Ui=Uj}\\ \Delta_{k,\sigma,\tau}^{(d_E)}&=\delta_{e_{\sigma(k)}e_k}\prod_{\ell=1}^k\delta_{e_{\sigma(\ell-1)}e^\prime_{\tau(\ell)}}
\delta_{e_{\ell-1}e^\prime_\ell},\label{DeltaEstate}
\end{align}
where $\{|e^{(\prime)}_i\rangle\}_{e_i=1}^{d_E}$, with $i=0,\ldots,k$, also summed over all its elements, is a set of environment bases and the $\Delta$ term is simply a monomial in $d_E$ with degree determined by $\sigma$ and $\tau$.

The case $k=0$ recovers $\avg(\Upsilon_{0:0})=\mbb{1}_S/d_S$ as expected, as no process occurs.

In the case of the superchannel $k=1$, we get Eq.~\eqref{avg state Ui=Uj superchannel}, which is quite different from the maximally mixed state that arises in the ergodic case, although it converges to it as $d_E\to\infty$. We notice that $\tr[\avg(\Upsilon_{1:0})]=1$ as expected and one may also verify the purity of this average state to be given by Eq.~\eqref{purity TI superchannel} which is in fact \emph{close} to that of the maximally mixed state for all $d_E$, $d_S$ and $1/d_S\leq\tr(\rho_S^2)\leq1$, and converges to it when $d_E\to\infty$ at a $\mc{O}(1/d_E)$ rate.

The detail of the calculation is as follows.

By definition,
\begin{align}&\avg(\Upsilon_{k:0})=\f{\tr_E}{d_S^k}\sum\int\limits_{U}U\FS_{\alpha_k\beta_k}
\cdots{U}\FS_{\alpha_1\beta_1}U\rho
{U}^\dg\FS_{\gamma_1\delta_1}^\dg{U}^\dg\cdots\FS_{\gamma_k\delta_k}^\dg{U}^\dg\,d\mu(U)\otimes|\beta_1\alpha_1\ldots\beta_k\alpha_k\rangle\!\langle\delta_1\gamma_1\ldots\delta_k\gamma_k|.\end{align}
We now consider directly decomposing the unitaries as $U=\sum{U}_{ab}|a\rangle\!\langle{b}|$ and $U^\dg=\sum{U}^*_{a^\prime{b}^\prime}|b^\prime\rangle\!\langle{a}^\prime|$ --notice that the $a$ and $b$ labels refer to the whole $ES$ space--, introducing an $E$ basis in the $\FS$ operators as $\FS_{ab}=\sum|ea\rangle\!\langle{eb}|$, and then evaluating the $k+1$ moments of $\mathrm{U}(d_Ed_S)$ by means of Eq.~\eqref{k moments of U},
\begin{align}\int\limits_U&U\FS_{\alpha_k\beta_k}
\cdots{U}\FS_{\alpha_1\beta_1}U\rho
{U}^\dg\FS_{\gamma_1\delta_1}^\dg{U}^\dg\cdots\FS_{\gamma_k\delta_k}^\dg{U}^\dg\,d\mu(U)\nonumber\\
&=\sum\int\limits_UU_{i_0j_0}\cdots{U}_{i_kj_k}
U^*_{i^\prime_0j^\prime_0}\cdots{U}^*_{i^\prime_kj^\prime_k}\,d\mu(U)\,|i_k\rangle\!\langle{j}_0|\rho|j^\prime_0\rangle\!\langle{i}^\prime_k|\,\prod_{\ell=1}^{k}\delta_{j_\ell(e\alpha)_\ell}\delta_{(e\beta)_\ell{i}_{\ell-1}}
\delta_{j^\prime_\ell(e^\prime\gamma)_\ell}\delta_{(e^\prime\delta)_\ell{i^\prime}_{\ell-1}}\nonumber\\
&=\sum\sum_{\sigma,\tau\in\mc{S}_{k+1}}\langle{j}_0|\rho|j^\prime_0\rangle\delta_{i_0i^\prime_{\sigma(0)}}\delta_{j_0j^\prime_{\tau(0)}}\prod_{\ell=1}^k
\delta_{i_\ell{i}^\prime_{\sigma(\ell)}}\delta_{j_\ell{j}^\prime_{\tau(\ell)}}\delta_{j_\ell(e\alpha)_\ell}\delta_{(e\beta)_\ell{i}_{\ell-1}}
\delta_{j^\prime_\ell(e^\prime\gamma)_\ell}\delta_{(e^\prime\delta)_\ell{i^\prime}_{\ell-1}}\,\mathrm{Wg}(\tau\sigma^{-1})\,|i_k\rangle\!\langle{i}^\prime_k|\nonumber\\
&=\sum\sum_{\sigma,\tau\in\mc{S}_{k+1}}\langle{j}^\prime_{\tau(0)}|\rho|
j^\prime_0\rangle\delta_{i_ki^\prime_{\sigma(k)}}
\prod_{\ell=1}^k\delta_{(e\beta)_\ell{i}^\prime_{\sigma(\ell-1)}}\delta_{(e^\prime\delta)_\ell{i^\prime}_{\ell-1}}
\delta_{(e\alpha)_\ell{j}^\prime_{\tau(\ell)}}\delta_{(e^\prime\gamma)_\ell{j}^\prime_\ell}\,\mathrm{Wg}(\tau\sigma^{-1})\,|i_k\rangle\!\langle{i}^\prime_k|,\end{align}
where here $\mc{S}_{k+1}$ denotes the symmetric group on $\{0,\ldots,k\}$, and which taking $i\to\epsilon\varsigma$ and $j\to\epsilon^\prime\varsigma^\prime$ to recover each $E$ and $S$ part explicitly, turns into
\begin{align}\int\limits_U&U\FS_{\alpha_k\beta_k}
\cdots{U}\FS_{\alpha_1\beta_1}U\rho
{U}^\dg\FS_{\gamma_1\delta_1}^\dg{U}^\dg\cdots\FS_{\gamma_k\delta_k}^\dg{U}^\dg\,d\mu(U)\nonumber\\
&=\sum\sum_{\sigma,\tau\in\mc{S}_{k+1}}\langle\epsilon^\prime_{\tau(0)}\varsigma^\prime_{\tau(0)}|\rho|
\epsilon^\prime_0\varsigma^\prime_0\rangle
\prod_{\ell=1}^k\delta_{\epsilon_{\sigma(\ell-1)}\epsilon^\prime_{\tau(\ell)}}\delta_{\epsilon_{\ell-1}\epsilon^\prime_\ell}
\delta_{\beta_\ell\varsigma_{\sigma(\ell-1)}}\delta_{\delta_\ell\varsigma_{\ell-1}}
\delta_{\alpha_\ell\varsigma^\prime_{\tau(\ell)}}\delta_{\gamma_\ell\varsigma^\prime_\ell}\,\mathrm{Wg}(\tau\sigma^{-1})\,|\epsilon_{\sigma(k)}\varsigma_{\sigma(k)}\rangle\!\langle\epsilon_k\varsigma_k|,\end{align}
and thus
\begin{align}&\avg(\Upsilon_{k:0})=\f{\sum}{d_S^k}\nonumber\\
&\sum_{\sigma,\tau\in\mc{S}_{k+1}}\hspace{-0.1in}\langle\epsilon^\prime_{\tau(0)}\varsigma^\prime_{\tau(0)}|\rho|
\epsilon^\prime_0\varsigma^\prime_0\rangle\mathrm{Wg}(\tau\sigma^{-1})
\delta_{\epsilon_{\sigma(k)}\epsilon_k}
\prod_{\ell=1}^k\delta_{\epsilon_{\sigma(\ell-1)}\epsilon^\prime_{\tau(\ell)}}\delta_{\epsilon_{\ell-1}\epsilon^\prime_\ell}
|\varsigma_{\sigma(k)}\varsigma_{\sigma(0)}\varsigma^\prime_{\tau(1)}\cdots\varsigma_{\sigma(k-1)}\varsigma^\prime_{\tau(k)}\rangle
\langle\varsigma_k\varsigma_0\varsigma^\prime_1\cdots\varsigma_{k-1}\varsigma^\prime_k|,\end{align}
as stated by Eq.\eqref{average state Ui=Uj}.

\subsubsection{Superchannel case}\label{appendix: average state Ui=Uj superchannel}
For the superchannel case, $k=1$, we have $\mc{S}_2=\{(0,1),(0)(1)\}$ where the elements are permutations stated in \emph{cycle} notation, representing the assignments $(0,1)=\begin{cases}0\to1\\1\to0\end{cases}$ and $(0)(1)=\boldsymbol{1}^2=\begin{cases}0\to0\\1\to1\end{cases}$, then (we write $\alpha,\beta,\gamma,\delta$ for $\varsigma_0,\varsigma_1,\varsigma_0^\prime,\varsigma_1^\prime$, respectively to ease the notation)
\begin{align}\avg(\Upsilon_{1:0})=\f{\sum}{d_S}\bigg\{&\left[\langle\delta|\rho_S|
\gamma\rangle|\alpha\beta\gamma\rangle\!\langle\beta\alpha\delta|+d_E^2|\beta\alpha\delta\rangle\!\langle\beta\alpha\delta|\right]\mathrm{Wg}[\boldsymbol{1}^2]+d_E\left[\langle\delta|\rho_S|
\gamma\rangle|\beta\alpha\gamma\rangle\!\langle\beta\alpha\delta|+|\alpha\beta\delta\rangle\!\langle\beta\alpha\delta|\right]\mathrm{Wg}[(0,1)]\bigg\},\end{align} so that with $\mathrm{Wg}[\boldsymbol{1}^2,d]=\f{1}{d^2-1}$ and $\mathrm{Wg}[(0,1),d]=\f{-1}{d(d^2-1)}$~\cite{GuMoments} we get
\begin{align}\avg(\Upsilon_{1:0})=\f{1}{d_E^2d_S^2-1}\bigg[\f{d_E^2}{d_S}\mbb1_{SAB}+\f{\swap}{d_S}\otimes\rho_{_S}^\intercal-\f{\swap}{d_S}\otimes\f{\mbb1_B}{d_S}-\f{\mbb1_{SA}}{d_S^2}\otimes\rho_{_S}^\intercal\bigg],\label{avg state Ui=Uj superchannel}\end{align} where $\swap=\sum_{i,j}|ij\rangle\!\langle{j}i|$ is the usual swap operator, and hence for the corresponding purity one may verify that
\begin{gather}\tr[\avg(\Upsilon_{1:0})^2]=\f{2}{(d_E^2d_S^2-1)^2}\left[\f{1}{d_S^3}+\tr(\rho_S^2)\f{d_S^2-d_S-1}{2d_S^2}-\f{d_E^2}{d_S}+\f{d_E^4d_S}{2}\right].\label{purity TI superchannel}\end{gather}

\subsubsection{Small subsystem limit}\label{appendix: average state Ui=Uj asympt}
When looking at the limit $d_E\to\infty$, the term that does not vanish is the one with $\sigma,\tau=\boldsymbol{1}^{k+1}$, i.e. with both permutations being identities, as these generate the most numerator powers in $d_E$ via the $\delta$ terms (correspondingly, $\Delta_{k,\sigma,\tau}^{(d_E)}$ in Eq.\eqref{DeltaEstate}) when summed over all $e_i$'s, i.e.
\begin{gather}\sum_{\substack{e^{(\prime)}_i=1\\e^\prime\neq{e}^\prime_0}}^{d_E}\delta_{\epsilon_k\epsilon_k}
\prod_{\ell=1}^k\delta_{\epsilon_{\ell-1}\epsilon^\prime_{\ell}}\delta_{\epsilon_{\ell-1}\epsilon^\prime_\ell}=d_E^{k+1},\end{gather}
all other permutations will vanish because of the $d_E$ powers in the denominator generated by the $\mathrm{Wg}$ function, the least powers produced by it are those when $\sigma\tau^{-1}=\mathbf{1}^{k+1}$ because $\#\boldsymbol{1}^n=\#[(1)(2)\cdots(n)]=n$, i.e. identity produces the greatest number of cycles, being the number of fixed points. Finally $\sum_{\epsilon^\prime_0\varsigma^\prime_0}\langle\epsilon^\prime_0\varsigma^\prime_0|\rho|
\epsilon^\prime_0\varsigma^\prime_0\rangle=\tr\rho=1$, and thus
\begin{gather}\avg(\Upsilon_{k:0})\sim\f{d_E^{k+1}}{d_S^k}
\mathrm{Wg}(\boldsymbol{1}^{k+1})\,\mbb{1}_{SA_1B_1\ldots{A}_kB_k}=\f{1}{d_S^{2k+1}}\,\mbb{1}_{SA_1B_1\ldots{A}_kB_k},\hspace{0.1in}\text{when}\,\,d_E\to\infty,\end{gather}
matching the average over an ergodic process.

\section{Proof of Main Result}\label{Appendix: Main Result}
\subsection{Preamble: Concentration of Measure (Levy's lemma)}\label{appendix: Concentration}
\subsubsection{Lipschitz functions}
A function $f:X\to{Y}$ between metric spaces $(X,\delta_{_X})$ and $(Y,\delta_{_Y})$ is $L$-Lipschitz if there is a real constant $\eta\geq0$, known as the Lipschitz constant, such that
\begin{gather}\delta_{_Y}\left(f(x_1),f(x_2)\right)\leq{L}\,\delta_{_X}(x_1,x_2).\end{gather}
Below we obtain the Lipschitz constant for the trace distance between a Choi process tensor and the maximally mixed state as a function of unitary operators. We consider both the time-independent ($U_i=U_j=U$ for all steps $i\neq{j}$) and the ergodic case ($U_i\neq{U}_j$  for all steps $i\neq{j}$).

\subsubsection{Levy's lemma}\label{app: Levy}
Let $M$ be a manifold with metric $\delta_{_M}$ and probability measure $\mu_{_M}$, and let $f:M\to\mathbb{R}$ be an $\eta$-Lipschitz function, then for all $\varepsilon>0$,
\begin{gather}\mathbb{P}_{x\sim\mu_{_M}}[f(x)\geq\mathbb{E}_x(f)+\varepsilon]\leq\alpha_{_M}(\varepsilon/\eta),\end{gather}
where $\alpha_{_M}$ is known as a concentration function for $M$, and is defined as (an upper bound on) the measure of the set of points in $M$ more than a distance $x$ from the minimal-boundary volume enclosing half of $M$. See~\cite{Ledoux, Philthy}.

\subsection{Main proof}\label{theorem proof}
We make use of Levy's lemma and the upper-bound $\mathbb{E}[\mc{N}]\leq\mc{B}_k(d_E,d_S)$ for the average non-Markovianity over the unitary group with the Haar measure, to prove our main result; specifically here we compute the corresponding Lipschitz constants and concentration functions for the function $D(\Upsilon,d_S^{-(2k+1)}\mbb1)$. We highlight that the evolution for $\Upsilon$ is sampled from the Haar measure, or equivalently one may replace $\mathbb{P}$ with something like $\mathbb{P}_{\Upsilon\sim\mu_{_{U_i}}}$.

\subsubsection{The Lipschitz constants}\label{appendix: Lipschitz}
We take $\tau:\mathrm{U}(d_Ed_S)\to\mathbb{R}$ defined by $\tau(U)=D(\Upsilon(U),\f{\mbb1}{d_S^{2k+1}})=\f{1}{2}\|\Upsilon(U)-\f{\mbb1}{d_S^{2k+1}}\|_1$ for the time-independent case, where we explicitly mean $\Upsilon(U)=\tr_E[\mc{U}_{k:0}\Theta\mc{U}_{k:0}^\dg]$ as in definition~\eqref{PT Choi state Def} with $\Theta=\rho\otimes\Psi^{\otimes{k}}$ a pure state, then for all $V\in\mathrm{U}(d_Ed_S)$,
\begin{align}|\tau({U})-\tau({V})|&=\left|D(\Upsilon(U),\f{\mbb1}{d_S^{2k+1}})-D(\Upsilon(V),\f{\mbb1}{d_S^{2k+1}})\right|\nonumber\\
&\leq{D}(\Upsilon(U),\Upsilon(V))\nonumber\\
&\leq{D}(\mc{U}_{k:0}\Theta\mc{U}_{k:0}^\dg,\mc{V}_{k:0}\Theta\mc{V}_{k:0}^\dg).\end{align}
Consider for now different labelings for the unitary at each time step. By the triangle inequality,
\begin{align}\|\mc{U}_{k:0}\Theta\mc{U}_{k:0}^\dg-\mc{V}_{k:0}\Theta\mc{V}_{k:0}^\dg\|_1&\leq\|\mc{U}_{k:0}\Theta\mc{U}_{k:0}^\dg-U_k\mc{S}_k\mc{V}_{k-1:0}\Theta\mc{V}_{k-1:0}^\dg\mc{S}_k^\dg{U}_k^\dg\|_1+\|U_k\mc{S}_k\mc{V}_{k-1:0}\Theta\mc{V}_{k-1:0}^\dg\mc{S}_k^\dg{U}_k^\dg-\mc{V}_{k:0}\Theta\mc{V}_{k:0}^\dg\|_1\nonumber\\
&=\|\mc{U}_{k-1:0}\Theta\mc{U}_{k-1:0}^\dg-\mc{V}_{k-1:0}\Theta\mc{V}_{k-1:0}^\dg\|_1+\|U_k\mc{S}_k\mc{V}_{k-1:0}\Theta\mc{V}_{k-1:0}^\dg\mc{S}_k^\dg{U}_k^\dg-\mc{V}_{k:0}\Theta\mc{V}_{k:0}^\dg\|_1,\label{LipIneq2}\end{align}
where it is clear by context that $U$ stands for $U\otimes\mbb1$.

Now we use Lemma 1 of~\cite{qspeedLipsch}, which states that
\begin{gather}\|A\sigma{A}^\dg-B\sigma{B}^\dg\|_1\leq2\|A-B\|_2\end{gather}
for two unitaries $A$, $B$ and any $\sigma$. For the simplest case with $k=1$, this gives
\begin{align}&\|U_1\mc{S}_1V_0\Theta{V}_0^\dg
\mc{S}_1^\dg{U}_1^\dg-V_1\mc{S}_1V_0\Theta{V}_0^\dg
\mc{S}_1^\dg{V}_1^\dg\|_1=\|\f{\sum}{d_S}(U_1\FS_{\alpha\beta}
V_0\rho{V}_0^\dg\FS_{\gamma\delta}U_1^\dg-
{V}_1\FS_{\alpha\beta}V_0\rho{V}_0^\dg\FS_{\gamma\delta}V_1^\dg)\otimes|\beta\alpha\rangle\!\langle\gamma\delta|\|_1\nonumber\\
&\leq\f{\sum_{\alpha\beta}}{d_S}\tr\sqrt{\sum_{\gamma\delta}(U_1\FS_{\alpha\beta}
V_0\rho{V}_0^\dg\FS_{\gamma\delta}U_1^\dg-
{V}_1\FS_{\alpha\beta}V_0\rho{V}_0^\dg\FS_{\gamma\delta}V_1^\dg)(U_1\FS_{\beta\alpha}
V_0\rho{V}_0^\dg\FS_{\delta\gamma}U_1^\dg-
{V}_1\FS_{\beta\alpha}V_0\rho{V}_0^\dg\FS_{\delta\gamma}V_1^\dg)}\nonumber\\
&\leq\f{\sum_{\alpha\beta}}{d_S}\|U_1\FS_{\alpha\beta}
V_0\rho{V}_0^\dg(\sum\FS_{\gamma\delta})U_1^\dg-
{V}_1\FS_{\alpha\beta}V_0\rho{V}_0^\dg(\sum\FS_{\gamma^\prime\delta^\prime})V_1^\dg\|_1\nonumber\\
&\leq\f{2}{d_S}\sum_{\alpha,\beta=1}^{d_S}\|U_1-V_1\|_2=2d_S\|U_1-V_1\|_2,\end{align}
and doing similarly, for the $i$-th step,
\begin{align}\|U_i\mc{S}_i\mc{V}_{i-1:0}\Theta\mc{V}_{i-1:0}^\dg\mc{S}_i^\dg{U}_i^\dg-\mc{V}_{i:0}\Theta\mc{V}_{i:0}^\dg\|_1
&\leq2d_S^i\|U_i-V_i\|.\end{align}
Thus bounding iteratively expression~\eqref{LipIneq2}, it follows that
\begin{align}\|\mc{U}_{k:0}\Theta\mc{U}_{k:0}^\dg-\mc{V}_{k:0}\Theta\mc{V}_{k:0}^\dg\|_1&\leq2\sum_{\ell=0}^kd_S^\ell\|U_\ell-V_\ell\|_2,\end{align}
finally giving
\begin{align}|\tau(U)-\tau(V)|&\leq\left(\f{d_S^{k+1}-1}{d_S-1}\right)\|U-V\|_2.\end{align}

On the other hand, for the ergodic case, let $\mathbb{U}(d)=\underbrace{\mathrm{U}(d)\times\cdots\times\mathrm{U}(d)}_{k+1\,\text{times}}$ be the $k+1$ Cartesian product space of $d$-dimensional unitary groups, then we define $\zeta:\mathbb{U}(d_Ed_S)\to\mathbb{R}$ by $\zeta(\vec{U})=D(\Upsilon(\vec{U}),\f{\mbb1}{d_S^{2k+1}})$ where now $\vec{U}=(U_0,\cdots,U_k)$. Similarly as before, we now have
\begin{gather}|\zeta(\vec{U})-\zeta(\vec{V})|\leq\sum_{\ell=0}^kd_S^\ell\|U_\ell-V_\ell\|_2,\end{gather}
and we may let the metric on $\mathbb{U}(d)$ be the 2-product metric $\delta_\mathbb{U}$ defined~\cite{deza2009encyclopedia} by
\begin{gather}\delta_\mathbb{U}(\vec{x},\vec{y})=\sqrt{\sum_{\ell=0}^k\|x_\ell-y_\ell\|_2^2},\end{gather}
which then satisfies
\begin{align}\sum_{\ell=0}^kd_S^\ell\|U_\ell-V_\ell\|_2&\leq\sum_{\ell=0}^kd_S^\ell\sqrt{\sum_{\ell^\prime=0}^k\|U_{\ell^\prime}-V_{\ell^\prime}\|_2^2}=\left(\f{d_S^{k+1}-1}{d_S-1}\right)\delta_\mathbb{U}(\vec{U},\vec{V}),
\end{align}
and thus we conclude that
\begin{gather}|\zeta(\vec{U})-\zeta(\vec{V})|\leq\left(\f{d_S^{k+1}-1}{d_S-1}\right)\delta_\mathbb{U}(\vec{U},\vec{V}),\end{gather}
so the (bound on) Lipschitz constants coincide with $\eta\leq\f{d_S^{k+1}-1}{d_S-1}$, which essentially behaves as $\mc{O}(d_S^k)$.

\subsubsection{The concentration function}
We now make use of a result related to the Gromov-Bishop inequality (see e.g. Theorem 7 in~\cite{Philthy}) stating that if $\mathrm{Ric}(M)\geq\mathrm{Ric}(\Sigma^n(R))=\f{n-1}{R^2}$ for an $n$-dimensional manifold $M$, where $\Sigma^n(R)$ is the $n$-dimensional sphere of radius $R$ and $\mathrm{Ric}(X)$ is the infimum of diagonal elements of the Ricci curvature tensor on $X$~\cite{Philthy}, then the respective concentration functions satisfy $\alpha_M(x)\leq\alpha_{\Sigma^n(R)}(x)\leq\exp\left[-\f{x^2(n-1)}{2R^2}\right]$~\cite{Ledoux}.

For the time-independent case, the corresponding manifold is the group manifold $\mathscr{U}$ of $\mathrm{U}(d)$ (where here $d=d_Ed_S$), which is diffeomorphic to $\mathrm{SU}(d)\times\Sigma^1(1)$ and thus has $\mathrm{Ric}(\mathscr{U})=d/2$ and $\dim(\mathscr{U})=d^2$. Then it follows that $\mathrm{Ric}(\mathscr{U})\geq\mathrm{Ric}(\Sigma^{d^2}(r))$ if $r^2\geq2(d^2-1)/d$ and hence, taking the minimal case,
\begin{gather}\alpha_{_\mathscr{U}}(\varepsilon/\eta)\leq\alpha_{\Sigma^{d^2}(r)}(\varepsilon/\eta)\leq\exp\left(-\f{\varepsilon^2d}{4\eta^2}\right).\end{gather}
For the ergodic case, the corresponding manifold is the group manifold $\mathbf{U}$ of the $k+1$ Cartesian product space $\mathrm{U}(d)\times\cdots\times\mathrm{U}(d)$. In this case $\mathrm{Ric}(\mathbf{U})=(k+1)d/2$ and $\dim(\mathbf{U})=(k+1)d^2$. Then it follows that $\mathrm{Ric}(\mathbf{U})\geq\mathrm{Ric}(\Sigma^{(k+1)d^2}(\mc{R}))$ if $\mc{R}^2\geq2[(k+1)d^2-1]/[d(k+1)]$ and hence, taking the minimal case,
\begin{gather}\alpha_{_\mathbf{U}}(\varepsilon/\eta)\leq\alpha_{\Sigma^{(k+1)d^2}(\mc{R})}(\varepsilon/\eta)\leq\exp\left(-\f{\varepsilon^2(k+1)d}{4\eta^2}\right).\end{gather}
The factor $d\eta^{-2}$ behaves as $\mc{O}(d_Ed_S^{-2k+1})$, so both concentration functions will be small whenever $d_E\gg{d}_S^{2k}$.

\section{Average purity}\label{sec:Purity Averages}
The purity of a quantum state is defined by the trace of its square and has an interpretation and relevance of its own in terms of correlations and/or indeterminacy of the state~\cite{Pasquale}. In terms of the Choi process tensor, it can offer insights about correlations and entanglement developed along a given process, as well as the \emph{noisiness} generated. Aside from a direct interpretation of the purity, we compute its average over the Haar measure for the process tensor in order to obtain explicit consequences on the average non-Markovianity.

\subsection{Ergodic case}\label{appendix: ergodic avg purity}
The average purity of the Choi process tensor for $k$-steps in the ergodic case is given by
\begin{gather}\avgdif[\tr(\Upsilon^{2})]=\f{d_E^2-1}{d_E(d_Ed_S+1)}\left(\f{d_E^2-1}{d_E^2d_S^2-1}\right)^k+\f{1}{d_E}.\end{gather}
We first notice that the simplest case $k=0$, i.e. with no process occurring, reproduces the now well known result $\avgdif[\tr(\Upsilon_{0:0}^{2})]=(d_E+d_S)/(d_Ed_S+1)$ found in~\cite{Lubkin, PageEntropy, LloydPagels, ScottCaves, Li, Giraud, Pasquale}.

The average purity Eq.~\eqref{average purity ergodic} decreases polynomially in $d_E$ and in the small subsystem limit satisfies
\begin{gather}\lim_{d_E\to\infty}\avgdif[\tr(\Upsilon^{2})]=\f{1}{d_S^{2k+1}},\end{gather}
thus leading to the purity of the maximally mixed state at a rate $\mc{O}(1/d_E)$. It is also manifestly decreasing in $k$ and in the long time limit satisfies
\begin{gather}\lim_{k\to\infty}\avgdif[\tr(\Upsilon^{2})]=\f{1}{d_E},\end{gather}
becoming independent of the dimension of the system. This limit corresponds strictly to a Choi process tensor with an infinite number of ancillary systems of dimension $d_S$ (making the purity bounded by zero from below) or more intuitively to a process after a large number of steps.

The detail of the calculation is as follows.

With $\Theta=\rho\otimes\Psi^{\otimes{k}}$, we first (following the approach in section 2 of~\cite{ZhangXiang}) write the trace as
\begin{align}\tr(\Upsilon_{k:0}^2)&=\tr\left[\left(\tr_E(\mc{U}_{k:0}\Theta\,\mc{U}^\dg_{k:0})\right)^2\right]\nonumber\\
&=\tr\left[\tr_E\left((\mbb{1}_E\otimes\Upsilon_{k:0})\,\mc{U}_{k:0}\Theta\,\mc{U}^\dg_{k:0}\right)\right]\nonumber\\
&=\tr\left[\Gamma_{_E}(\mc{U}_{k:0}\Theta\,\mc{U}^\dg_{k:0})\,\mc{U}_{k:0}\Theta\,\mc{U}^\dg_{k:0}\right]\nonumber\\
&=\tr\left[\Theta\,\mc{U}_{k:0}^\dg\Gamma_{_E}(\mc{U}_{k:0}\Theta\,\mc{U}^\dg_{k:0})\,\mc{U}_{k:0}\right],\end{align}
where $\Gamma_{_E}(\cdot)\equiv\mbb{1}_E\otimes\tr_E(\cdot)$ is a CP map and can be expressed as an operator-sum, i.e. for any $\mc{X}$ acting on $ESA_1B_1\cdots{A}_kB_k$ we have $\Gamma_{_E}(\mc{X})=\sum_{i,j=1}^{d_E}(K_{ij}\otimes\mbb1)\mc{X}(K_{ij}\otimes\mbb1)^\dg$ where the identities are on the ancillary system and $K_{ij}=|i\rangle\!\langle{j}|\otimes\mbb{1}_{S}$ are the ($ES$ system) Kraus operators with $\{|i\rangle\}_{i=1}^{d_E}$ a given basis for $E$. And so we need to compute the integrals \begin{align}\Omega_{k:0}\equiv\int\limits_{U_k\cdots\,U_0}\mc{U}_{k:0}^\dg\Gamma_E(\mc{U}_{k:0}\Theta\,\mc{U}^\dg_{k:0})\,\mc{U}_{k:0}\,d\mu(U_0)\cdots\,d\mu(U_k),\label{full purity integrals}\end{align}
and in fact these are really in the $ES$ part only, which has the form
\begin{gather}\omega_{k:0}^{(\rho)}\equiv\int\limits_UU_0^\dg
\FS_{x_1\alpha_1}U_1^\dg\cdots{\FS}_{x_k\alpha_k}
U_k^\dg{K}_{\tilde\imath\tilde\jmath}U_k\FS_{\beta_kx_k}\cdots{U_1}\FS_{\beta_1x_1}U_0\rho
{U}_0^\dg\FS_{y_1\lambda_1}U_1^\dg\cdots\FS_{y_k\lambda_k}U_k^\dg{K}^\dg_{\tilde\imath\tilde\jmath}{U}_k\FS_{\theta_ky_k}\cdots{U}_1\FS_{\theta_1y_1}U_0,\label{ES part purity integ}\end{gather}
where we dropped the Haar measure symbols just to economize notation. Summation over repeated indices $x,y$ and $i,j$ on the Kraus operators is implied in this expression and the ancillary part takes the form
\begin{align}\Omega_{k:0}^{(AB)}&=(|\alpha_1\rangle\!\langle\beta_1|\otimes\mbb1_{B_1}\otimes\cdots\otimes|\alpha_k\rangle\!\langle\beta_k|\otimes\mbb1_{B_k})\Psi^{\otimes\,k}
(|\lambda_1\rangle\!\langle\theta_1|\otimes\mbb1_{B_1}\otimes\cdots\otimes|\lambda_k\rangle\!\langle\theta_k|\otimes\mbb1_{B_k}).\end{align}
We may then evaluate the integrals from $U_0$ to $U_k$ with the result
\begin{gather}\int\limits_{\mathrm{U}(d)}U^\dg{A}UX{U}^\dg{B}U\,d\mu(U)=\f{d\tr(AB)-\tr(A)\tr(B)}{d(d^2-1)}\tr(X)\mbb1+\f{d\tr(A)\tr(B)-\tr(AB)}{d(d^2-1)}X,\label{intmp}\end{gather}
also, here $d=d_Ed_S$, then we have
\begin{align}&\omega_{k:0}^{(\rho)}=\f{1}{d(d^2-1)}\nonumber\\
&\bigg(\delta_{y_1x_1}\delta_{x_1y_1}\hspace{-0.15in}\int\limits_{U_k\cdots{U}_1}
\hspace{-0.15in}\tr\left[\FS_{\theta_1\alpha_1}U_1^\dg\cdots{\FS}_{x_k\alpha_k}
U_k^\dg{K}_{ij}U_k\FS_{\beta_kx_k}\hspace{-0.035in}\cdots{U}_1\FS_{\beta_1\lambda_1}U_1^\dg\cdots\FS_{y_k\lambda_k}U_k^\dg{K}^\dg_{ij}{U}_k\FS_{\theta_ky_k}\hspace{-0.035in}\cdots{U}_1\right](d\mbb1-\rho)\nonumber\\
&+\delta_{x_1x_1}\delta_{y_1y_1}\hspace*{-0.15in}\int\limits_{U_k\cdots{U}_1}\hspace*{-0.15in}\tr\left[\FS_{\beta_1\alpha_1}U_1^\dg\cdots{\FS}_{x_k\alpha_k}
U_k^\dg{K}_{ij}U_k\FS_{\beta_kx_k}\cdots{U}_1\right]\tr\left[\FS_{\theta_1\lambda_1}U_1^\dg\cdots\FS_{y_k\lambda_k}U_k^\dg{K}^\dg_{ij}{U}_k\FS_{\theta_ky_k}\cdots{U}_1\right](d\rho-\mbb1)\bigg)\nonumber\\
&=\f{1}{d(d^2-1)}\bigg[d_S\langle{e}_1\alpha_1|\omega_{k:1}^{(\FS_{\beta_1\lambda_1})}|e_1\theta_1\rangle(d\mbb1-\rho)+d_S^2\langle{e}_1\alpha_1|\omega_{k:1}^{(|e_1\beta_1\rangle\!\langle{e}^\prime_1\lambda_1|)}|e^\prime_1\theta_1\rangle(d\rho-\mbb1)\bigg],\end{align}
where sum is implied over $e_i$'s, and so on, only relying on Eq.~\eqref{intmp}. First, let us state that
\begin{align}\omega_{k:i-1}^{(X)}&=\f{1}{d(d^2-1)}\bigg[d_S\langle{e}_i\alpha_i|\omega_{k:i}^{(\FS_{\beta_i\lambda_i})}|e_i\theta_i\rangle(d\tr(X)\mbb1-X)+d_S^2\langle{e}_i\alpha_i|\omega_{k:i}^{(|e_i\beta_i\rangle\!\langle{e}^\prime_i\lambda_i|)}|e^\prime_i\theta_i\rangle(dX-\tr(X)\mbb1)\bigg],\end{align}
for $k>i-1\geq0$. Then we get
\begin{align}&\omega_{k:0}^{(\rho)}=\f{1}{d^2(d^2-1)^2}\bigg\{\nonumber\\
&d_S\langle{e}_2\alpha_2|\omega_{k:2}^{(\FS_{\beta_2\lambda_2})}|e_2\theta_2\rangle\bigg[d_S(dd_E^2\delta_{\alpha_1\theta_1}\delta_{\beta_1\lambda_1}-d_E\delta_{\alpha_1\beta_1}\delta_{\lambda_1\theta_1})(d\mbb1-\rho)+d_S^2(dd_E\delta_{\alpha_1\theta_1}\delta_{\beta_1\lambda_1}-d_E^2\delta_{\alpha_1\beta_1}\delta_{\lambda_1\theta_1})(d\rho-\mbb1)\bigg]\nonumber\\
&\hspace{-0.05in}+d_S^2\langle{e}_2\alpha_2|\omega_{k:2}^{(|e_2\beta_2\rangle\!\langle{e}_2^\prime\lambda_2|)}|e_2^\prime\theta_2\rangle\bigg[d_S(dd_E\delta_{\alpha_1\beta_1}\delta_{\lambda_1\theta_1}\hspace*{-0.05in}-d_E^2\delta_{\alpha_1\theta_1}\delta_{\beta_1\lambda_1})(d\mbb1-\rho)+d_S^2(dd_E^2\delta_{\alpha_1\beta_1}\delta_{\lambda_1\theta_1}\hspace*{-0.05in}-d_E\delta_{\alpha_1\theta_1}\delta_{\beta_1\lambda_1})(d\rho-\mbb1)\bigg]\bigg\}.\label{int1_k:0}\end{align}
Before carrying on, let us notice that the case $\omega_{k:k}$ is special because here one evaluates the terms on the Kraus operators $\tr(K_{ij}K_{ij}^\dg)$ and $\tr(K_{ij})\tr(K_{ij}^\dg)$ summed over their indices, which leaves
\begin{gather}\langle{e}_k\alpha_k|\omega_{k:k}^{(\FS_{\lambda_k\beta_k})}|e_k\theta_k\rangle=\f{1}{d(d^2-1)}[dd_E^2(dd_E\delta_{\alpha_k\theta_k}\delta_{\beta_k\lambda_k}
-\delta_{\alpha_k\beta_k}\delta_{\lambda_k\theta_k})+d^2d_E(d_S\delta_{\alpha_k\beta_k}\delta_{\lambda_k\theta_k}
-\delta_{\alpha_k\theta_k}\delta_{\beta_k\lambda_k})],
\end{gather}
\begin{gather}\langle{e}_k\alpha_k|\omega_{k:k}^{(|e_k\beta_k\rangle\!\langle{e}^\prime_k\lambda_k|)}|e^\prime_k\theta_k\rangle=\f{1}{d(d^2-1)}[dd_E^3(d_S\delta_{\alpha_k\theta_k}\delta_{\beta_k\lambda_k}-\delta_{\alpha_k\beta_k}\delta_{\lambda_k\theta_k})+
d^2(dd_E\delta_{\alpha_k\beta_k}\delta_{\lambda_k\theta_k}-\delta_{\alpha_k\theta_k}\delta_{\beta_k\lambda_k})].
\end{gather}
Notice that we can get rid of the Kronecker deltas easily in the full $\Omega_{k:0}$ when summing over Greek indices, as $\delta_{\alpha\theta}\delta_{\beta\lambda}$ give rise to maximally mixed states in the ancillas while terms $\delta_{\alpha\beta}\delta_{\lambda\theta}$ give rise to identities, and there are only deltas of this kind. Thus we may simply assign $\delta_{\alpha\theta}\delta_{\beta\lambda}\to1/d_S$ and $\delta_{\alpha\beta}\delta_{\lambda\theta}\to1$ when plugging the corresponding expressions in $\Omega_{2:0}$\textsuperscript{\footnote{ In particular this implies that the terms $d_S\delta_{\alpha\theta}\delta_{\beta\lambda}-\delta_{\alpha\beta}\delta_{\lambda\theta}$ and $dd_E\delta_{\alpha\beta}\delta_{\lambda\theta}-d_E^2\delta_{\alpha\theta}\delta_{\beta\lambda}$ won't contribute to the average purity.}}. Furthermore, a direct consequence of this is that $\Omega_{k:0}$ will be a linear combination of $2^{k+1}$ outer products between $\mbb1$, $\rho$ and $\Psi$, implying that (as $\rho$ is pure, $\Psi$ is idempotent with trace one and the trace of an outer product is the product of traces) the average purity $\avgdif[\tr(\Upsilon^2_{k:0})]$ will be a sum of the scalar terms in $\omega^{(\rho)}_{k:0}$ together with an extra overall factor of $d-1$. Let us denote by
\begin{gather}A=d_E(d_E^2-1),\quad
B=d_E^2-d_E^2=0,\quad
C=d_E^2(d_S-1/d_S),\quad
D=d_E(dd_E-1/d_S),\end{gather}
the terms that appear in expression~\eqref{int1_k:0} after having taken $\delta_{\alpha\theta}\delta_{\beta\lambda}\to1/d_S$ and 
$\delta_{\alpha\beta}\delta_{\lambda\theta}\to1$. Doing similarly for the $\omega_{k:k}$ case,
\begin{gather}\mc{A}=dd_E^2(d_E^2+d_S^2-2),\quad
\mc{B}=dd_E(d^2-1),\end{gather}
we find that\textsuperscript{\footnote{ A way to deduce this is to sub-label each factor $A, B,\cdots$ by the $i$ index of the term ${\langle{e}_i\alpha_i|\omega^{X_i}|e_i^{(\prime)}\theta_i\rangle}$ that they come from and then substituting recursively in Eq.~\eqref{int1_k:0} as if one were already evaluating the whole purity. The vanishing of $B$ simplifies this process a great deal.}}
\begin{align}\avgdif[\tr(\Upsilon_{k:0}^{2})]&=\frac{d_S^k(d-1)}{d^{k+1}(d^2-1)^{k+1}}\left[\mc{A}\,A^{k-1}+d_S\mc{B}\left(CA^{k-2}+C\sum_{i=1}^{k-2}d_S^iD^iA^{k-i-2}+d_S^{k-1}D^{k-1}\right)\right],\end{align}
where the terms with a $C$ factor are taken as zero for $k=1$ and as $C$ for $k=2$. We may then simplify this expression by means of the geometric series to the right hand side of Eq.~\eqref{average purity ergodic}.

\subsection{Time-independent case}\label{appendix: purity Ui=Uj}
We make use of Eq.~\eqref{k moments of U} with the dimension argument of the Weingarten function taken implicitly as $d_Ed_S$. The average purity of the Choi process tensor for $k$ steps in a time-independent process can be expressed with implicit sum over repeated indices, as
\begin{gather}\avg[\tr(\Upsilon^2)]=\f{1}{d_S^{2k}}\sum_{\sigma,\tau\in\mc{S}_{2k+2}}\mathrm{Wg}(\tau\sigma^{-1})\,\rho_{_{\tau(0);k+1}}\rho_{_{\tau(k+1);0}}\tilde\Delta_{k,\sigma,\tau}^{(d_E)}\tilde{\tilde\Delta}_{k,\sigma,\tau}^{(d_S)},\end{gather}
where the notation in $\rho$ is as in Eq.~\eqref{phi notation Ui=Uj}, and where now the symmetric group is on $\{0,\ldots,2k+1\}$; we also make the definitions in Eq.~\eqref{TIpuritydef} for $\tilde\Delta^{(d_E)}$, $\tilde{\tilde\Delta}^{(d_S)}$, which are monomials in $d_E$ and $d_S$ correspondingly with degree determined by $\sigma$, $\tau$. The number of elements in the symmetric group scale as $(2k+2)!$ and expressions like Eq.~\eqref{average time independent purity} are generally non-trivial (see also e.g.~\cite{Pasquale}); however, we can gain valuable insights from it by looking at their asymptotics. Overall, numerical sampling suggest that this average purity remains quite close to that of the ergodic case.

For $k=0$, with no process, we may also verify that we recover $\avg[\tr(\Upsilon_{0:0}^{2})]=(d_E+d_S)/(d_Ed_S+1)$  as expected~\cite{Lubkin, PageEntropy, LloydPagels, ScottCaves, Li, Giraud, Pasquale}.

In the small subsystem limit, as detailed below in~\ref{appendix: purity Ui=Uj asympt}, we obtain
\begin{gather}\lim_{d_E\to\infty}\avg[\tr(\Upsilon^2)]=\f{1}{d_S^{2k+1}},\end{gather}
corresponding to the purity in the ergodic case in this limit. Similarly, in the long time limit, we obtain in~\ref{appendix: purity Ui=Uj asympt time},
\begin{gather}\lim_{k\to\infty}\avg[\tr(\Upsilon^{2})]=\f{1}{d_E},\end{gather}
and the average purity in this limit also coincides with that of ergodic processes.

The detail of the calculation is as follows.

We may write the integral in the $ES$ part, as in Eq.\eqref{ES part purity integ} now with same unitary (throughout we take sums over repeated indices and omit the Haar measure symbol),
\begin{gather}\omega_{k:0}^{(\rho)}\equiv\int\limits_UU^\dg
\FS_{x_1\alpha_1}U^\dg\cdots{\FS}_{x_k\alpha_k}
U^\dg{K}_{\tilde\imath\tilde\jmath}U\FS_{\beta_kx_k}\cdots{U}\FS_{\beta_1x_1}U\rho
{U}^\dg\FS_{y_1\lambda_1}U^\dg\cdots\FS_{y_k\lambda_k}U^\dg{K}^\dg_{\tilde\imath\tilde\jmath}{U}\FS_{\theta_ky_k}\cdots{U}\FS_{\theta_1y_1}U.\label{ES integral k step}\end{gather}
We decompose the unitaries in the whole $ES$ space as $U=\sum{U}_{ab}|a\rangle\!\langle{b}|$ and $U^\dg=\sum{U}^*_{a^\prime{b}^\prime}|b^\prime\rangle\!\langle{a}^\prime|$ (i.e. the $a$, $b$ labels here refer to the whole $ES$ space) and we enumerate the labels of the ones to the left of $\rho$ in Eq.~\eqref{ES integral k step} from $0$ to $k$ (priming, ${}^\prime$, the adjoint ones) and the remaining from $k+1$ to $2k+1$, we also do this in increasing order for the adjoint unitaries and decreasing order for the original unitaries so that they match the order of the $\FS$ operators, i.e. the unitary components will originally appear as $U^*_{i^\prime_0j^\prime_0}\cdots{U}^*_{i^\prime_kj^\prime_k}U_{i_kj_k}\cdots{U_{i_0j_0}}
U^*_{i^\prime_{k+1}j^\prime_{k+1}}\cdots{U}^*_{i^\prime_{2k+1}j^\prime_{2k+1}}U_{i^\prime_{2k+1}j^\prime_{2k+1}}\cdots
{U_{i^\prime_{k+1}j^\prime_{k+1}}}$, and we then rearrange them to the form of Eq.~\eqref{k moments of U} just keeping track of the correct order in the operator part, for which we introduce sets of $E$-basis into each $\FS$ operator as $\FS_{ab}=\sum|ea\rangle\!\langle{eb}|$ labeled by $e$ for the ones to the left of $\rho$ in Eq.~\eqref{ES integral k step}, priming the ones between adjoint unitaries, and by $\epsilon$ the corresponding ones to the right, also priming the ones between unitaries.

This then leads to
\begin{align}\omega_{k:0}^{(\rho)}&=\int\limits_UU_{i_0j_0}\cdots{U}_{i_{2k+1}j_{2k+1}}U^*_{i^\prime_0j^\prime_0}
\cdots{U}^*_{i^\prime_{2k+1}j^\prime_{2k+1}}\,\langle{i}^\prime_k|K_{\tilde\imath\tilde\jmath}|i_k\rangle\!\langle{j}_0|\rho|j^\prime_{k+1}\rangle\!\langle{i}^\prime_{2k+1}|K^\dg_{\tilde\imath\tilde\jmath}|i_{2k+1}\rangle|j_0^\prime\rangle\!\langle{j}_{k+1}|\nonumber\\
&\hspace{2in}\prod_{\ell=1}^k\delta_{i_{\ell-1}(ex)_\ell}\delta_{j_\ell(e\beta)_\ell}\delta_{i_{\ell+k}(\epsilon{y})_\ell}\delta_{j_{\ell+k+1}(\epsilon\theta)_\ell}\delta_{i^\prime_{\ell-1}(e^\prime{x})_\ell}\delta_{j_\ell^\prime(e^\prime\alpha)_\ell}
\delta_{i^\prime_{\ell+k}(\epsilon^\prime{y})_\ell}\delta_{j_{\ell+k+1}^\prime(\epsilon^\prime\lambda)_\ell}\nonumber\\
&=\langle{i}^\prime_k|K_{\tilde\imath\tilde\jmath}|i_k\rangle\!\langle{j}_0|\rho|j^\prime_{k+1}\rangle\!\langle{i}^\prime_{2k+1}|K^\dg_{\tilde\imath\tilde\jmath}|i_{2k+1}\rangle|j_0^\prime\rangle\!\langle{j}_{k+1}|\sum_{\sigma,\tau\in\mc{S}_{2k+2}}\mathrm{Wg}(\tau\sigma^{-1})\prod_{n=0}^{2k+1}\delta_{i_ni^\prime_{\sigma(n)}}\delta_{j_nj^\prime_{\tau(n)}}\nonumber\\
&\hspace{1.5in}
\prod_{\ell=1}^k\delta_{i_{\ell-1}(ex)_\ell}\delta_{j_\ell(e\beta)_\ell}\delta_{i_{\ell+k}(\epsilon{y})_\ell}\delta_{j_{\ell+k+1}(\epsilon\theta)_\ell}\delta_{i^\prime_{\ell-1}(e^\prime{x})_\ell}\delta_{j_\ell^\prime(e^\prime\alpha)_\ell}
\delta_{i^\prime_{\ell+k}(\epsilon^\prime{y})_\ell}\delta_{j_{\ell+k+1}^\prime(\epsilon^\prime\lambda)_\ell}\nonumber\\
&=\sum_{\sigma,\tau\in\mc{S}_{2k+2}}\mathrm{Wg}(\tau\sigma^{-1})\langle{i}^\prime_k|K_{\tilde\imath\tilde\jmath}|i^\prime_{\sigma(k)}\rangle\!\langle{j}^\prime_{\tau(0)}|\rho|j^\prime_{k+1}\rangle\!\langle{i}^\prime_{2k+1}|K^\dg_{\tilde\imath\tilde\jmath}|i^\prime_{\sigma(2k+1)}\rangle|j_0^\prime\rangle\!\langle{j}^\prime_{\tau(k+1)}|\nonumber\\
&\hspace{1in}\prod_{\ell=1}^k\delta_{i^\prime_{\sigma(\ell-1)}(ex)_\ell}\delta_{j^\prime_{\tau(\ell)}(e\beta)_\ell}\delta_{i^\prime_{\sigma(\ell+k)}(\epsilon{y})_\ell}
\delta_{j^\prime_{\tau(\ell+k+1)}(\epsilon\theta)_\ell}\delta_{i^\prime_{\ell-1}(e^\prime{x})_\ell}\delta_{j_\ell^\prime(e^\prime\alpha)_\ell}
\delta_{i^\prime_{\ell+k}(\epsilon^\prime{y})_\ell}\delta_{j_{\ell+k+1}^\prime(\epsilon^\prime\lambda)_\ell}.\end{align}

We now replace $i^\prime\to\varepsilon\varsigma$ and $j^\prime\to\varepsilon^\prime\varsigma^\prime$ in order to split the $E$ and $S$ parts explicitly, leaving
\begin{align}&\omega_{k:0}^{(\rho)}\nonumber\\
&=\hspace{-0.1in}\sum_{\sigma,\tau\in\mc{S}_{2k+2}}\hspace{-0.15in}\mathrm{Wg}(\tau\sigma^{-1})\langle\varepsilon_k\varsigma_k|K_{\tilde\imath\tilde\jmath}|\varepsilon_{\sigma(k)}\varsigma_{\sigma(k)}\rangle\!\langle\varepsilon^\prime_{\tau(0)}\varsigma^\prime_{\tau(0)}|\rho|\varepsilon^\prime_{k+1}\varsigma^\prime_{k+1}\rangle\!\langle\varepsilon_{2k+1}\varsigma_{2k+1}|K^\dg_{\tilde\imath\tilde\jmath}|\varepsilon_{\sigma(2k+1)}\varsigma_{\sigma(2k+1)}\rangle|\varepsilon^\prime_0\varsigma^\prime_0\rangle\!\langle\varepsilon^\prime_{\tau(k+1)}\varsigma^\prime_{\tau(k+1)}|\nonumber\\
&\hspace{0.5in}\prod_{\ell=1}^k\delta_{(\varepsilon\varsigma)_{\sigma(\ell-1)}(ex)_\ell}\delta_{(\varepsilon^\prime\varsigma^\prime)_{\tau(\ell)}(e\beta)_\ell}\delta_{(\varepsilon\varsigma)_{\sigma(\ell+k)}(\epsilon{y})_\ell}
\delta_{(\varepsilon^\prime\varsigma^\prime)_{\tau(\ell+k+1)}(\epsilon\theta)_\ell}\delta_{(\varepsilon\varsigma)_{\ell-1}(e^\prime{x})_\ell}\delta_{(\varepsilon^\prime\varsigma^\prime)_\ell(e^\prime\alpha)_\ell}
\delta_{(\varepsilon\varsigma)_{\ell+k}(\epsilon^\prime{y})_\ell}\delta_{(\varepsilon^\prime\varsigma^\prime)_{\ell+k+1}(\epsilon^\prime\lambda)_\ell}\nonumber\\
&=\hspace{-0.1in}\sum_{\sigma,\tau\in\mc{S}_{2k+2}}\hspace{-0.15in}\mathrm{Wg}(\tau\sigma^{-1})\langle\varepsilon^\prime_{\tau(0)}\varsigma^\prime_{\tau(0)}|\rho|\varepsilon^\prime_{k+1}\varsigma^\prime_{k+1}\rangle
|\varepsilon^\prime_0\varsigma^\prime_0\rangle\!\langle\varepsilon^\prime_{\tau(k+1)}\varsigma^\prime_{\tau(k+1)}|
\delta_{\varepsilon_k\varepsilon_{\sigma(2k+1)}}\delta_{\varepsilon_{\sigma(k)}\varepsilon_{2k+1}}\nonumber\\
&\hspace{0.5in}\prod_{\ell=1}^k\delta_{\varepsilon_{\sigma(\ell-1)}\varepsilon^\prime_{\tau(\ell)}}
\delta_{\varepsilon_{\sigma(\ell+k)}\varepsilon^\prime_{\tau(\ell+k+1)}}\delta_{\varepsilon_{\ell-1}\varepsilon^\prime_\ell}
\delta_{\varepsilon_{\ell+k}\varepsilon^\prime_{\ell+k+1}}
\delta_{\varsigma^\prime_\ell\alpha_\ell}\delta_{\varsigma^\prime_{\tau(\ell)}\beta_\ell}\delta_{\varsigma^\prime_{\tau(\ell+k+1)}\theta_\ell}
\delta_{\varsigma^\prime_{\ell+k+1}\lambda_\ell}\prod_{n=0}^{2k+1}\delta_{\varsigma_{\sigma(n)}\varsigma_n}.\end{align}

The ancillary part is simply $\Omega_{k:0}^{(AB)}=d_S^{-k}|\alpha_1\beta_1\ldots\alpha_k\beta_k\rangle\!\langle\theta_1\lambda_1\ldots\theta_k\lambda_k|$, so we get for the analogue of Eq.~\eqref{full purity integrals},
\begin{align}&\Omega_{k:0}=\omega_{k:0}^{(\rho)}\otimes\Omega_{k:0}^{(AB)}\nonumber\\
&=\f{1}{d_S^k}\sum_{\sigma,\tau\in\mc{S}_{2k+2}}\hspace{-0.15in}\mathrm{Wg}(\tau\sigma^{-1})
\delta_{\varepsilon_k\varepsilon_{\sigma(2k+1)}}\delta_{\varepsilon_{\sigma(k)}\varepsilon_{2k+1}}
|\varepsilon^\prime_0\varsigma^\prime_0\varsigma^\prime_1\varsigma^\prime_{\tau(1)}\cdots
\varsigma^\prime_k\varsigma^\prime_{\tau(k)}\rangle
\langle\varepsilon^\prime_{\tau(k+1)}\varsigma^\prime_{\tau(k+1)}\varsigma^\prime_{\tau(k+2)}\varsigma^\prime_{k+2}\cdots
\varsigma^\prime_{\tau(2k+1)}\varsigma^\prime_{2k+1}|\nonumber\\
&\hspace{0.5in}\langle\varepsilon^\prime_{\tau(0)}\varsigma^\prime_{\tau(0)}|\rho|\varepsilon^\prime_{k+1}\varsigma^\prime_{k+1}\rangle\prod_{n=0}^{2k+1}\delta_{\varsigma_{\sigma(n)}\varsigma_n}\prod_{\ell=1}^k
\delta_{\varepsilon_{\sigma(\ell-1)}\varepsilon^\prime_{\tau(\ell)}}
\delta_{\varepsilon_{\sigma(\ell+k)}\varepsilon^\prime_{\tau(\ell+k+1)}}\delta_{\varepsilon_{\ell-1}\varepsilon^\prime_\ell}
\delta_{\varepsilon_{\ell+k}\varepsilon^\prime_{\ell+k+1}},\end{align}
and thus
\begin{align}\avg[\tr(\Upsilon_{k:0}^2)]&=\tr[(\rho\otimes\Psi^{\otimes{k}})\,\Omega_{k:0}]\nonumber\\
&=\f{1}{d_S^{2k}}\sum_{\sigma,\tau\in\mc{S}_{2k+2}}\hspace{-0.15in}\mathrm{Wg}(\tau\sigma^{-1})\langle\varepsilon^\prime_{\tau(0)}\varsigma^\prime_{\tau(0)}|\rho|\varepsilon^\prime_{k+1}\varsigma^\prime_{k+1}\rangle\!\langle\varepsilon^\prime_{\tau(k+1)}\varsigma^\prime_{\tau(k+1)}|\rho|\varepsilon^\prime_{0}\varsigma^\prime_{0}\rangle
\delta_{\varepsilon_k\varepsilon_{\sigma(2k+1)}}\delta_{\varepsilon_{2k+1}\varepsilon_{\sigma(k)}}\nonumber\\
&\hspace{0.5in}\prod_{\substack{\ell=1\\\ell\neq{k+1}}}^{2k+1}
\delta_{\varepsilon_{\ell-1}\varepsilon^\prime_\ell}\delta_{\varepsilon_{\sigma(\ell-1)}\varepsilon^\prime_{\tau(\ell)}}
\delta_{\varsigma^\prime_\ell\varsigma^\prime_{\tau(\ell)}}\prod_{n=0}^{2k+1}\delta_{\varsigma_{\sigma(n)}\varsigma_n},\end{align}
as stated in Eq.~\eqref{average time independent purity}, where we make the definitions
\begin{equation}\tilde\Delta_{k,\sigma,\tau}^{(d_E)}=\delta_{e_ke_{\sigma(2k+1)}}\delta_{e_{2k+1}e_{\sigma(k)}}
\prod_{\substack{\ell=1\\\ell\neq{k+1}}}^{2k+1}
\delta_{e_{\ell-1}e^\prime_\ell}\delta_{e_{\sigma(\ell-1)}e^\prime_{\tau(\ell)}},\quad
\tilde{\tilde\Delta}_{k,\sigma,\tau}^{(d_S)}=\prod_{\substack{\ell=1\\\ell\neq{k+1}}}^{2k+1}
\delta_{s^\prime_\ell{s}^\prime_{\tau(\ell)}}\prod_{n=0}^{2k+1}\delta_{s_ns_{\sigma(n)}}\label{TIpuritydef}\end{equation}

\subsubsection{Small subsystem limit}\label{appendix: purity Ui=Uj asympt}
In the $d_E\to\infty$ limit the only term that remains is the one with
\begin{gather}\sigma=\tau=(0,k+1)(1,k+2)\cdots(k,2k+1),\end{gather}
as expressed in cycle notation, which simply means $\sigma(0)=k+1=\tau(0)$, $\sigma(k+1)=0=\tau(k+1)$, and so on. This then leads to the correct limit, as
\begin{align}\tilde{\tilde\Delta}_{k,\sigma,\tau}^{(d_S)}\to\sum_{\substack{\varsigma^{(\prime)}_i=1\\\varsigma^\prime\neq\{\varsigma^\prime_0,\varsigma^\prime_{k+1}\}}}^{d_S}
\prod_{\substack{\ell=1\\\ell\neq{k+1}}}^{2k+1}\delta_{\varsigma^\prime_\ell\varsigma^\prime_{\tau(\ell)}}
\prod_{n=0}^{2k+1}\delta_{\varsigma_{\sigma(n)}\varsigma_n}
&={d_S}^{2k+1},\\
\tilde\Delta_{k,\sigma,\tau}^{(d_E)}\to\sum_{\substack{\varepsilon^{(\prime)}_i=1\\\varepsilon^\prime\neq\{\varepsilon^\prime_0,\varepsilon^\prime_{k+1}\}}}^{d_E}
\delta_{\varepsilon_k\varepsilon_{\sigma(2k+1)}}\delta_{\varepsilon_{2k+1}\varepsilon_{\sigma(k)}}
\prod_{\substack{\ell=1\\\ell\neq{k+1}}}^{2k+1}
\delta_{\varepsilon_{\ell-1}\varepsilon^\prime_\ell}\delta_{\varepsilon_{\sigma(\ell-1)}\varepsilon^\prime_{\tau(\ell)}}
&={d_E}^{2k+2},\end{align}
while keeping $\sigma\tau^{-1}=\mathbf{1}^{2k+2}$, i.e. with the argument in the $\mathrm{Wg}$ function an identity, which generates the least denominator powers because $\#\boldsymbol{1}^n=\#[(1)(2)\cdots(n)]=n$, i.e. the identity generates the most number of cycles. When considering other permutations for $\sigma$ and $\tau$ we see that the $d_E$ powers in the numerator can only decrease while those in the denominator can only increase (as the identity is the permutation that generates the most cycles) so indeed in the large subsystem limit this is the only pair of permutations that survive; the terms in $\rho$ yield traces when summed over basis vectors and thus give factors of one,
\begin{gather}\sum_{\varepsilon^\prime_{k+1},\varsigma^\prime_{k+1}}\langle\varepsilon^\prime_{k+1}\varsigma^\prime_{k+1}|\rho|\varepsilon^\prime_{k+1}\varsigma^\prime_{k+1}\rangle=\sum_{\varepsilon^\prime_0,\varsigma^\prime_0}\langle\varepsilon^\prime_0\varsigma^\prime_0|\rho|\varepsilon^\prime_{0}\varsigma^\prime_{0}\rangle=\tr\rho=1.\end{gather}
 This leads then to the conclusion
\begin{align}\avg[\tr(\Upsilon_{k:0}^2)]&\sim\f{d_S^{2k+1}d_E^{2k+2}}{d_S^{2k}}\mathrm{Wg}(\mathbf{1}^{2k+2})=d_Sd_E^{2k+2}\f{1}{(d_Ed_S)^{2k+2}}=\f{1}{d_S^{2k+1}},\hspace{0.1in}\text{when}\,\,d_E\to\infty.\end{align}

\subsubsection{Long time limit}\label{appendix: purity Ui=Uj asympt time}
For the case when $k\to\infty$, the $\mathrm{Wg}$ function behaves dominantly as in the small subsystem limit and the only pair of permutations that identify all dimension powers in the numerator without dependence of $k$ are identities $\sigma=\tau=\mathbf{1}^{2k+2}$. In this case
\begin{align}\sum_{\substack{\varsigma^{(\prime)}_i=1\\\varsigma^\prime\neq\{\varsigma^\prime_0,\varsigma^\prime_{k+1}\}}}^{d_S}
\prod_{\substack{\ell=1\\\ell\neq{k+1}}}^{2k+1}\delta_{\varsigma^\prime_\ell\varsigma^\prime_\ell}
\prod_{n=0}^{2k+1}\delta_{\varsigma_n\varsigma_n}
&={d_S}^{4k+2},\\
\sum_{\substack{\varepsilon^{(\prime)}_i=1\\\varepsilon^\prime\neq\{\varepsilon^\prime_0,\varepsilon^\prime_{k+1}\}}}^{d_E}
\delta_{\varepsilon_k\varepsilon_{2k+1}}\delta_{\varepsilon_{2k+1}\varepsilon_{k}}
\prod_{\substack{\ell=1\\\ell\neq{k+1}}}^{2k+1}
\delta_{\varepsilon_{\ell-1}\varepsilon^\prime_\ell}\delta_{\varepsilon_{\ell-1}\varepsilon^\prime_{\ell}}
&={d_E}^{2k+1},\end{align}
giving
\begin{align}\avg[\tr(\Upsilon_{k:0}^2)]&\sim\f{d_S^{4k+2}d_E^{2k+1}}{d_S^{2k}}\tr(\rho^2)\mathrm{Wg}(\mathbf{1}^{2k+2})=\f{d_S^{2k+2}d_E^{2k+1}}{(d_Ed_S)^{2k+2}}=\f{1}{d_E},\hspace{0.1in}\text{when}\,\,k\to\infty,\end{align}
as $\tr(\rho^2)=1$ by assumption.


\begin{thebibliography}{58}%
\makeatletter
\providecommand \@ifxundefined [1]{%
 \@ifx{#1\undefined}
}%
\providecommand \@ifnum [1]{%
 \ifnum #1\expandafter \@firstoftwo
 \else \expandafter \@secondoftwo
 \fi
}%
\providecommand \@ifx [1]{%
 \ifx #1\expandafter \@firstoftwo
 \else \expandafter \@secondoftwo
 \fi
}%
\providecommand \natexlab [1]{#1}%
\providecommand \enquote  [1]{``#1''}%
\providecommand \bibnamefont  [1]{#1}%
\providecommand \bibfnamefont [1]{#1}%
\providecommand \citenamefont [1]{#1}%
\providecommand \href@noop [0]{\@secondoftwo}%
\providecommand \href [0]{\begingroup \@sanitize@url \@href}%
\providecommand \@href[1]{\@@startlink{#1}\@@href}%
\providecommand \@@href[1]{\endgroup#1\@@endlink}%
\providecommand \@sanitize@url [0]{\catcode `\\12\catcode `\$12\catcode
  `\&12\catcode `\#12\catcode `\^12\catcode `\_12\catcode `\%12\relax}%
\providecommand \@@startlink[1]{}%
\providecommand \@@endlink[0]{}%
\providecommand \url  [0]{\begingroup\@sanitize@url \@url }%
\providecommand \@url [1]{\endgroup\@href {#1}{\urlprefix }}%
\providecommand \urlprefix  [0]{URL }%
\providecommand \Eprint [0]{\href }%
\providecommand \doibase [0]{http://dx.doi.org/}%
\providecommand \selectlanguage [0]{\@gobble}%
\providecommand \bibinfo  [0]{\@secondoftwo}%
\providecommand \bibfield  [0]{\@secondoftwo}%
\providecommand \translation [1]{[#1]}%
\providecommand \BibitemOpen [0]{}%
\providecommand \bibitemStop [0]{}%
\providecommand \bibitemNoStop [0]{.\EOS\space}%
\providecommand \EOS [0]{\spacefactor3000\relax}%
\providecommand \BibitemShut  [1]{\csname bibitem#1\endcsname}%
\let\auto@bib@innerbib\@empty
\bibitem [{\citenamefont {Gogolin}\ and\ \citenamefont
  {Eisert}(2016)}]{Gogolin}%
  \BibitemOpen
  \bibfield  {author} {\bibinfo {author} {\bibfnamefont {C.}~\bibnamefont
  {Gogolin}}\ and\ \bibinfo {author} {\bibfnamefont {J.}~\bibnamefont
  {Eisert}},\ }\bibfield  {title} {\enquote {\bibinfo {title} {Equilibration,
  thermalisation, and the emergence of statistical mechanics in closed quantum
  systems},}\ }\href {\doibase 10.1088/0034-4885/79/5/056001} {\bibfield
  {journal} {\bibinfo  {journal} {Rep. Prog. Phys.}\ }\textbf {\bibinfo
  {volume} {79}},\ \bibinfo {pages} {056001} (\bibinfo {year}
  {2016})}\BibitemShut {NoStop}%
\bibitem [{\citenamefont {Goold}\ \emph {et~al.}(2016)\citenamefont {Goold},
  \citenamefont {Huber}, \citenamefont {Riera}, \citenamefont {del Rio},\ and\
  \citenamefont {Skrzypczyk}}]{Goold}%
  \BibitemOpen
  \bibfield  {author} {\bibinfo {author} {\bibfnamefont {J.}~\bibnamefont
  {Goold}}, \bibinfo {author} {\bibfnamefont {M.}~\bibnamefont {Huber}},
  \bibinfo {author} {\bibfnamefont {A.}~\bibnamefont {Riera}}, \bibinfo
  {author} {\bibfnamefont {L.}~\bibnamefont {del Rio}}, \ and\ \bibinfo
  {author} {\bibfnamefont {P.}~\bibnamefont {Skrzypczyk}},\ }\bibfield  {title}
  {\enquote {\bibinfo {title} {The role of quantum information in
  thermodynamics--a topical review},}\ }\href {\doibase
  10.1088/1751-8113/49/14/143001} {\bibfield  {journal} {\bibinfo  {journal}
  {J. Phys. A: Math. Theor.}\ }\textbf {\bibinfo {volume} {49}},\ \bibinfo
  {pages} {143001} (\bibinfo {year} {2016})}\BibitemShut {NoStop}%
\bibitem [{\citenamefont {Garc\'{\i}a-Pintos}\ \emph
  {et~al.}(2017)\citenamefont {Garc\'{\i}a-Pintos}, \citenamefont {Linden},
  \citenamefont {Malabarba}, \citenamefont {Short},\ and\ \citenamefont
  {Winter}}]{GarciaPintos}%
  \BibitemOpen
  \bibfield  {author} {\bibinfo {author} {\bibfnamefont {L.~P.}\ \bibnamefont
  {Garc\'{\i}a-Pintos}}, \bibinfo {author} {\bibfnamefont {N.}~\bibnamefont
  {Linden}}, \bibinfo {author} {\bibfnamefont {A.~S.~L.}\ \bibnamefont
  {Malabarba}}, \bibinfo {author} {\bibfnamefont {A.~J.}\ \bibnamefont
  {Short}}, \ and\ \bibinfo {author} {\bibfnamefont {A.}~\bibnamefont
  {Winter}},\ }\bibfield  {title} {\enquote {\bibinfo {title} {Equilibration
  time scales of physically relevant observables},}\ }\href {\doibase
  10.1103/PhysRevX.7.031027} {\bibfield  {journal} {\bibinfo  {journal} {Phys.
  Rev. X}\ }\textbf {\bibinfo {volume} {7}},\ \bibinfo {pages} {031027}
  (\bibinfo {year} {2017})}\BibitemShut {NoStop}%
\bibitem [{\citenamefont {Srednicki}(1999)}]{Srednicki1999}%
  \BibitemOpen
  \bibfield  {author} {\bibinfo {author} {\bibfnamefont {M.}~\bibnamefont
  {Srednicki}},\ }\bibfield  {title} {\enquote {\bibinfo {title} {The approach
  to thermal equilibrium in quantized chaotic systems},}\ }\href {\doibase
  10.1088/0305-4470/32/7/007} {\bibfield  {journal} {\bibinfo  {journal} {J.
  Phys. A: Math. Gen.}\ }\textbf {\bibinfo {volume} {32}},\ \bibinfo {pages}
  {1163} (\bibinfo {year} {1999})}\BibitemShut {NoStop}%
\bibitem [{\citenamefont {Pollock}\ \emph
  {et~al.}(2018{\natexlab{a}})\citenamefont {Pollock}, \citenamefont
  {Rodr\'{\i}guez-Rosario}, \citenamefont {Frauenheim}, \citenamefont
  {Paternostro},\ and\ \citenamefont {Modi}}]{processtensor2}%
  \BibitemOpen
  \bibfield  {author} {\bibinfo {author} {\bibfnamefont {F.~A.}\ \bibnamefont
  {Pollock}}, \bibinfo {author} {\bibfnamefont {C.}~\bibnamefont
  {Rodr\'{\i}guez-Rosario}}, \bibinfo {author} {\bibfnamefont {T.}~\bibnamefont
  {Frauenheim}}, \bibinfo {author} {\bibfnamefont {M.}~\bibnamefont
  {Paternostro}}, \ and\ \bibinfo {author} {\bibfnamefont {K.}~\bibnamefont
  {Modi}},\ }\bibfield  {title} {\enquote {\bibinfo {title} {Operational markov
  condition for quantum processes},}\ }\href {\doibase
  10.1103/PhysRevLett.120.040405} {\bibfield  {journal} {\bibinfo  {journal}
  {Phys. Rev. Lett.}\ }\textbf {\bibinfo {volume} {120}},\ \bibinfo {pages}
  {040405} (\bibinfo {year} {2018}{\natexlab{a}})}\BibitemShut {NoStop}%
\bibitem [{\citenamefont {Arenz}\ \emph {et~al.}(2015)\citenamefont {Arenz},
  \citenamefont {Hillier}, \citenamefont {Fraas},\ and\ \citenamefont
  {Burgarth}}]{PhysRevA.92.022102}%
  \BibitemOpen
  \bibfield  {author} {\bibinfo {author} {\bibfnamefont {C.}~\bibnamefont
  {Arenz}}, \bibinfo {author} {\bibfnamefont {R.}~\bibnamefont {Hillier}},
  \bibinfo {author} {\bibfnamefont {M.}~\bibnamefont {Fraas}}, \ and\ \bibinfo
  {author} {\bibfnamefont {D.}~\bibnamefont {Burgarth}},\ }\bibfield  {title}
  {\enquote {\bibinfo {title} {Distinguishing decoherence from alternative
  quantum theories by dynamical decoupling},}\ }\href {\doibase
  10.1103/PhysRevA.92.022102} {\bibfield  {journal} {\bibinfo  {journal} {Phys.
  Rev. A}\ }\textbf {\bibinfo {volume} {92}},\ \bibinfo {pages} {022102}
  (\bibinfo {year} {2015})}\BibitemShut {NoStop}%
\bibitem [{\citenamefont {Schlosshauer}(2007)}]{Schlosshauer2007}%
  \BibitemOpen
  \bibfield  {author} {\bibinfo {author} {\bibfnamefont {M.~A.}\ \bibnamefont
  {Schlosshauer}},\ }\href {\doibase 10.1007/978-3-540-35775-9} {\emph
  {\bibinfo {title} {Decoherence and the Quantum-To-Classical Transition}}}\
  (\bibinfo  {publisher} {Springer Berlin Heidelberg},\ \bibinfo {year}
  {2007})\BibitemShut {NoStop}%
\bibitem [{\citenamefont {de~Vega}\ and\ \citenamefont
  {Alonso}(2017)}]{VegaAlonso}%
  \BibitemOpen
  \bibfield  {author} {\bibinfo {author} {\bibfnamefont {I.}~\bibnamefont
  {de~Vega}}\ and\ \bibinfo {author} {\bibfnamefont {D.}~\bibnamefont
  {Alonso}},\ }\bibfield  {title} {\enquote {\bibinfo {title} {Dynamics of
  non-{Markovian} open quantum systems},}\ }\href {\doibase
  10.1103/RevModPhys.89.015001} {\bibfield  {journal} {\bibinfo  {journal}
  {Rev. Mod. Phys.}\ }\textbf {\bibinfo {volume} {89}},\ \bibinfo {pages}
  {015001} (\bibinfo {year} {2017})}\BibitemShut {NoStop}%
\bibitem [{\citenamefont {Ciccarello}(2017)}]{CollisionModels}%
  \BibitemOpen
  \bibfield  {author} {\bibinfo {author} {\bibfnamefont {F.}~\bibnamefont
  {Ciccarello}},\ }\bibfield  {title} {\enquote {\bibinfo {title} {Collision
  models in quantum optics},}\ }\href {\doibase 10.1515/qmetro-2017-0007}
  {\bibfield  {journal} {\bibinfo  {journal} {Quantum Meas. Quantum Metrol.}\
  }\textbf {\bibinfo {volume} {4}},\ \bibinfo {pages} {53} (\bibinfo {year}
  {2017})}\BibitemShut {NoStop}%
\bibitem [{\citenamefont {Spohn}\ and\ \citenamefont {Lebowitz}(2007)}]{spohn}%
  \BibitemOpen
  \bibfield  {author} {\bibinfo {author} {\bibfnamefont {H.}~\bibnamefont
  {Spohn}}\ and\ \bibinfo {author} {\bibfnamefont {J.~L.}\ \bibnamefont
  {Lebowitz}},\ }\bibfield  {title} {\enquote {\bibinfo {title} {Irreversible
  thermodynamics for quantum systems weakly coupled to thermal reservoirs},}\
  }in\ \href {\doibase 10.1002/9780470142578.ch2} {\emph {\bibinfo {booktitle}
  {Advances in Chemical Physics}}},\ \bibinfo {editor} {edited by\ \bibinfo
  {editor} {\bibfnamefont {S.~A.}\ \bibnamefont {Rice}}}\ (\bibinfo
  {publisher} {John Wiley \& Sons, Ltd},\ \bibinfo {year} {2007})\ pp.\
  \bibinfo {pages} {109--142}\BibitemShut {NoStop}%
\bibitem [{\citenamefont {Ledoux}(2001)}]{Ledoux}%
  \BibitemOpen
  \bibfield  {author} {\bibinfo {author} {\bibfnamefont {M.}~\bibnamefont
  {Ledoux}},\ }\href {https://bookstore.ams.org/surv-89-s} {\emph {\bibinfo
  {title} {The Concentration of Measure Phenomenon}}},\ Mathematical surveys
  and monographs\ (\bibinfo  {publisher} {American Mathematical Society},\
  \bibinfo {year} {2001})\BibitemShut {NoStop}%
\bibitem [{\citenamefont {Milman}\ and\ \citenamefont
  {Schechtman}(1986)}]{milman}%
  \BibitemOpen
  \bibfield  {author} {\bibinfo {author} {\bibfnamefont {V.~D.}\ \bibnamefont
  {Milman}}\ and\ \bibinfo {author} {\bibfnamefont {G.}~\bibnamefont
  {Schechtman}},\ }\href {\doibase 10.1007/978-3-540-38822-7} {\emph {\bibinfo
  {title} {Asymptotic Theory of Finite Dimensional Normed Spaces}}},\ \bibinfo
  {series} {Lecture Notes in Mathematics}\ No.\ \bibinfo {number} {1200}\
  (\bibinfo  {publisher} {Springer-Verlag},\ \bibinfo {year}
  {1986})\BibitemShut {NoStop}%
\bibitem [{\citenamefont {Boucheron}\ \emph {et~al.}(2013)\citenamefont
  {Boucheron}, \citenamefont {Lugosi},\ and\ \citenamefont
  {Massart}}]{boucheron2013concentration}%
  \BibitemOpen
  \bibfield  {author} {\bibinfo {author} {\bibfnamefont {S.}~\bibnamefont
  {Boucheron}}, \bibinfo {author} {\bibfnamefont {G.}~\bibnamefont {Lugosi}}, \
  and\ \bibinfo {author} {\bibfnamefont {P.}~\bibnamefont {Massart}},\ }\href
  {\doibase 10.1093/acprof:oso/9780199535255.001.0001} {\emph {\bibinfo {title}
  {Concentration Inequalities: A Nonasymptotic Theory of Independence}}}\
  (\bibinfo  {publisher} {OUP Oxford},\ \bibinfo {year} {2013})\BibitemShut
  {NoStop}%
\bibitem [{\citenamefont {Popescu}\ \emph {et~al.}(2006)\citenamefont
  {Popescu}, \citenamefont {Short},\ and\ \citenamefont
  {Winter}}]{PopescuWinter}%
  \BibitemOpen
  \bibfield  {author} {\bibinfo {author} {\bibfnamefont {S.}~\bibnamefont
  {Popescu}}, \bibinfo {author} {\bibfnamefont {A.~J.}\ \bibnamefont {Short}},
  \ and\ \bibinfo {author} {\bibfnamefont {A.}~\bibnamefont {Winter}},\
  }\bibfield  {title} {\enquote {\bibinfo {title} {Entanglement and the
  foundations of statistical mechanics},}\ }\href {\doibase 10.1038/nphys444}
  {\bibfield  {journal} {\bibinfo  {journal} {Nat. Phys.}\ }\textbf {\bibinfo
  {volume} {2}},\ \bibinfo {pages} {754} (\bibinfo {year} {2006})}\BibitemShut
  {NoStop}%
\bibitem [{\citenamefont {Li}(2013)}]{Li}%
  \BibitemOpen
  \bibfield  {author} {\bibinfo {author} {\bibfnamefont {Y.}~\bibnamefont
  {Li}},\ }\href@noop {} {\enquote {\bibinfo {title} {Seminar 6 of selected
  topics in mathematical physics: Quantum information theory},}\ }\bibinfo
  {howpublished} {\url{http://www.mpmueller.net/seminar/talk6.pdf}} (\bibinfo
  {year} {2013})\BibitemShut {NoStop}%
\bibitem [{\citenamefont {Masanes}\ \emph {et~al.}(2013)\citenamefont
  {Masanes}, \citenamefont {Roncaglia},\ and\ \citenamefont
  {Ac\'{\i}n}}]{Masanes}%
  \BibitemOpen
  \bibfield  {author} {\bibinfo {author} {\bibfnamefont {L.}~\bibnamefont
  {Masanes}}, \bibinfo {author} {\bibfnamefont {A.~J.}\ \bibnamefont
  {Roncaglia}}, \ and\ \bibinfo {author} {\bibfnamefont {A.}~\bibnamefont
  {Ac\'{\i}n}},\ }\bibfield  {title} {\enquote {\bibinfo {title} {Complexity of
  energy eigenstates as a mechanism for equilibration},}\ }\href {\doibase
  10.1103/PhysRevE.87.032137} {\bibfield  {journal} {\bibinfo  {journal} {Phys.
  Rev. E}\ }\textbf {\bibinfo {volume} {87}},\ \bibinfo {pages} {032137}
  (\bibinfo {year} {2013})}\BibitemShut {NoStop}%
\bibitem [{\citenamefont {Chiribella}\ \emph {et~al.}(2008)\citenamefont
  {Chiribella}, \citenamefont {D'Ariano},\ and\ \citenamefont
  {Perinotti}}]{PhysRevLett.101.060401}%
  \BibitemOpen
  \bibfield  {author} {\bibinfo {author} {\bibfnamefont {G.}~\bibnamefont
  {Chiribella}}, \bibinfo {author} {\bibfnamefont {G.~M.}\ \bibnamefont
  {D'Ariano}}, \ and\ \bibinfo {author} {\bibfnamefont {P.}~\bibnamefont
  {Perinotti}},\ }\bibfield  {title} {\enquote {\bibinfo {title} {Quantum
  circuit architecture},}\ }\href {\doibase 10.1103/PhysRevLett.101.060401}
  {\bibfield  {journal} {\bibinfo  {journal} {Phys. Rev. Lett.}\ }\textbf
  {\bibinfo {volume} {101}},\ \bibinfo {pages} {060401} (\bibinfo {year}
  {2008})}\BibitemShut {NoStop}%
\bibitem [{\citenamefont {Chiribella}\ \emph {et~al.}(2009)\citenamefont
  {Chiribella}, \citenamefont {D'Ariano},\ and\ \citenamefont
  {Perinotti}}]{PhysRevA.80.022339}%
  \BibitemOpen
  \bibfield  {author} {\bibinfo {author} {\bibfnamefont {G.}~\bibnamefont
  {Chiribella}}, \bibinfo {author} {\bibfnamefont {G.~M.}\ \bibnamefont
  {D'Ariano}}, \ and\ \bibinfo {author} {\bibfnamefont {P.}~\bibnamefont
  {Perinotti}},\ }\bibfield  {title} {\enquote {\bibinfo {title} {Theoretical
  framework for quantum networks},}\ }\href {\doibase
  10.1103/PhysRevA.80.022339} {\bibfield  {journal} {\bibinfo  {journal} {Phys.
  Rev. A}\ }\textbf {\bibinfo {volume} {80}},\ \bibinfo {pages} {022339}
  (\bibinfo {year} {2009})}\BibitemShut {NoStop}%
\bibitem [{\citenamefont {Costa}\ and\ \citenamefont {Shrapnel}(2016)}]{Costa}%
  \BibitemOpen
  \bibfield  {author} {\bibinfo {author} {\bibfnamefont {F.}~\bibnamefont
  {Costa}}\ and\ \bibinfo {author} {\bibfnamefont {S.}~\bibnamefont
  {Shrapnel}},\ }\bibfield  {title} {\enquote {\bibinfo {title} {Quantum causal
  modelling},}\ }\href {\doibase 10.1088/1367-2630/18/6/063032} {\bibfield
  {journal} {\bibinfo  {journal} {New J. Phys.}\ }\textbf {\bibinfo {volume}
  {18}},\ \bibinfo {pages} {063032} (\bibinfo {year} {2016})}\BibitemShut
  {NoStop}%
\bibitem [{\citenamefont {Modi}(2012)}]{Modi}%
  \BibitemOpen
  \bibfield  {author} {\bibinfo {author} {\bibfnamefont {K.}~\bibnamefont
  {Modi}},\ }\bibfield  {title} {\enquote {\bibinfo {title} {Operational
  approach to open dynamics and quantifying initial correlations},}\ }\href
  {\doibase 10.1038/srep00581} {\bibfield  {journal} {\bibinfo  {journal} {Sci.
  Rep.}\ }\textbf {\bibinfo {volume} {2}},\ \bibinfo {pages} {581} (\bibinfo
  {year} {2012})}\BibitemShut {NoStop}%
\bibitem [{\citenamefont {Pollock}\ \emph
  {et~al.}(2018{\natexlab{b}})\citenamefont {Pollock}, \citenamefont
  {Rodr\'{\i}guez-Rosario}, \citenamefont {Frauenheim}, \citenamefont
  {Paternostro},\ and\ \citenamefont {Modi}}]{processtensor}%
  \BibitemOpen
  \bibfield  {author} {\bibinfo {author} {\bibfnamefont {F.~A.}\ \bibnamefont
  {Pollock}}, \bibinfo {author} {\bibfnamefont {C.}~\bibnamefont
  {Rodr\'{\i}guez-Rosario}}, \bibinfo {author} {\bibfnamefont {T.}~\bibnamefont
  {Frauenheim}}, \bibinfo {author} {\bibfnamefont {M.}~\bibnamefont
  {Paternostro}}, \ and\ \bibinfo {author} {\bibfnamefont {K.}~\bibnamefont
  {Modi}},\ }\bibfield  {title} {\enquote {\bibinfo {title} {Non-{Markovian}
  quantum processes: Complete framework and efficient characterization},}\
  }\href {\doibase 10.1103/PhysRevA.97.012127} {\bibfield  {journal} {\bibinfo
  {journal} {Phys. Rev. A}\ }\textbf {\bibinfo {volume} {97}},\ \bibinfo
  {pages} {012127} (\bibinfo {year} {2018}{\natexlab{b}})}\BibitemShut
  {NoStop}%
\bibitem [{\citenamefont {Breuer}\ \emph {et~al.}(2016)\citenamefont {Breuer},
  \citenamefont {Laine}, \citenamefont {Piilo},\ and\ \citenamefont
  {Vacchini}}]{Breuer2015}%
  \BibitemOpen
  \bibfield  {author} {\bibinfo {author} {\bibfnamefont {H.-P.}\ \bibnamefont
  {Breuer}}, \bibinfo {author} {\bibfnamefont {E.-M.}\ \bibnamefont {Laine}},
  \bibinfo {author} {\bibfnamefont {J.}~\bibnamefont {Piilo}}, \ and\ \bibinfo
  {author} {\bibfnamefont {B.}~\bibnamefont {Vacchini}},\ }\bibfield  {title}
  {\enquote {\bibinfo {title} {Colloquium: Non-{Markovian} dynamics in open
  quantum systems},}\ }\href {\doibase 10.1103/RevModPhys.88.021002} {\bibfield
   {journal} {\bibinfo  {journal} {Rev. Mod. Phys.}\ }\textbf {\bibinfo
  {volume} {88}},\ \bibinfo {pages} {021002} (\bibinfo {year}
  {2016})}\BibitemShut {NoStop}%
\bibitem [{\citenamefont {Choi}(1975)}]{Choi1975}%
  \BibitemOpen
  \bibfield  {author} {\bibinfo {author} {\bibfnamefont {M.-D.}\ \bibnamefont
  {Choi}},\ }\bibfield  {title} {\enquote {\bibinfo {title} {Completely
  positive linear maps on complex matrices},}\ }\href {\doibase
  10.1016/0024-3795(75)90075-0} {\bibfield  {journal} {\bibinfo  {journal}
  {Linear Algebra Its Appl.}\ }\textbf {\bibinfo {volume} {10}},\ \bibinfo
  {pages} {285 } (\bibinfo {year} {1975})}\BibitemShut {NoStop}%
\bibitem [{\citenamefont {Jamio\l{}kowski}(1972)}]{Jamiol1972}%
  \BibitemOpen
  \bibfield  {author} {\bibinfo {author} {\bibfnamefont {A.}~\bibnamefont
  {Jamio\l{}kowski}},\ }\bibfield  {title} {\enquote {\bibinfo {title} {Linear
  transformations which preserve trace and positive semidefiniteness of
  operators},}\ }\href {\doibase 10.1016/0034-4877(72)90011-0} {\bibfield
  {journal} {\bibinfo  {journal} {Rep. Math. Phys.}\ }\textbf {\bibinfo
  {volume} {3}},\ \bibinfo {pages} {275 } (\bibinfo {year} {1972})}\BibitemShut
  {NoStop}%
\bibitem [{\citenamefont {Milz}\ \emph
  {et~al.}(2017{\natexlab{a}})\citenamefont {Milz}, \citenamefont {Pollock},\
  and\ \citenamefont {Modi}}]{Simon2017}%
  \BibitemOpen
  \bibfield  {author} {\bibinfo {author} {\bibfnamefont {S.}~\bibnamefont
  {Milz}}, \bibinfo {author} {\bibfnamefont {F.~A.}\ \bibnamefont {Pollock}}, \
  and\ \bibinfo {author} {\bibfnamefont {K.}~\bibnamefont {Modi}},\ }\bibfield
  {title} {\enquote {\bibinfo {title} {An introduction to operational quantum
  dynamics},}\ }\href {\doibase 10.1142/S1230161217400169} {\bibfield
  {journal} {\bibinfo  {journal} {Open Syst. Inf. Dyn.}\ }\textbf {\bibinfo
  {volume} {24}},\ \bibinfo {pages} {1740016} (\bibinfo {year}
  {2017}{\natexlab{a}})}\BibitemShut {NoStop}%
\bibitem [{\citenamefont {Short}\ and\ \citenamefont
  {Farrelly}(2012)}]{ShortFarrelly}%
  \BibitemOpen
  \bibfield  {author} {\bibinfo {author} {\bibfnamefont {A.~J.}\ \bibnamefont
  {Short}}\ and\ \bibinfo {author} {\bibfnamefont {T.~C.}\ \bibnamefont
  {Farrelly}},\ }\bibfield  {title} {\enquote {\bibinfo {title} {Quantum
  equilibration in finite time},}\ }\href {\doibase
  10.1088/1367-2630/14/1/013063} {\bibfield  {journal} {\bibinfo  {journal}
  {New J. Phys.}\ }\textbf {\bibinfo {volume} {14}},\ \bibinfo {pages} {013063}
  (\bibinfo {year} {2012})}\BibitemShut {NoStop}%
\bibitem [{\citenamefont {Gogolin}\ \emph {et~al.}(2011)\citenamefont
  {Gogolin}, \citenamefont {M\"uller},\ and\ \citenamefont
  {Eisert}}]{PhysRevLett.106.040401}%
  \BibitemOpen
  \bibfield  {author} {\bibinfo {author} {\bibfnamefont {C.}~\bibnamefont
  {Gogolin}}, \bibinfo {author} {\bibfnamefont {M.~P.}\ \bibnamefont
  {M\"uller}}, \ and\ \bibinfo {author} {\bibfnamefont {J.}~\bibnamefont
  {Eisert}},\ }\bibfield  {title} {\enquote {\bibinfo {title} {Absence of
  thermalization in nonintegrable systems},}\ }\href {\doibase
  10.1103/PhysRevLett.106.040401} {\bibfield  {journal} {\bibinfo  {journal}
  {Phys. Rev. Lett.}\ }\textbf {\bibinfo {volume} {106}},\ \bibinfo {pages}
  {040401} (\bibinfo {year} {2011})}\BibitemShut {NoStop}%
\bibitem [{\citenamefont {Gilchrist}\ \emph {et~al.}(2005)\citenamefont
  {Gilchrist}, \citenamefont {Langford},\ and\ \citenamefont
  {Nielsen}}]{PhysRevA.71.062310}%
  \BibitemOpen
  \bibfield  {author} {\bibinfo {author} {\bibfnamefont {A.}~\bibnamefont
  {Gilchrist}}, \bibinfo {author} {\bibfnamefont {N.~K.}\ \bibnamefont
  {Langford}}, \ and\ \bibinfo {author} {\bibfnamefont {M.~A.}\ \bibnamefont
  {Nielsen}},\ }\bibfield  {title} {\enquote {\bibinfo {title} {Distance
  measures to compare real and ideal quantum processes},}\ }\href {\doibase
  10.1103/PhysRevA.71.062310} {\bibfield  {journal} {\bibinfo  {journal} {Phys.
  Rev. A}\ }\textbf {\bibinfo {volume} {71}},\ \bibinfo {pages} {062310}
  (\bibinfo {year} {2005})}\BibitemShut {NoStop}%
\bibitem [{\citenamefont {Zhang}(2014)}]{ZhangMInt}%
  \BibitemOpen
  \bibfield  {author} {\bibinfo {author} {\bibfnamefont {L.}~\bibnamefont
  {Zhang}},\ }\bibfield  {title} {\enquote {\bibinfo {title} {Matrix integrals
  over unitary groups: An application of {Schur-Weyl} duality},}\ }\href
  {https://arxiv.org/abs/1408.3782} {\bibfield  {journal} {\bibinfo  {journal}
  {arXiv:1408.3782}\ } (\bibinfo {year} {2014})}\BibitemShut {NoStop}%
\bibitem [{\citenamefont {Gu}(2013)}]{GuMoments}%
  \BibitemOpen
  \bibfield  {author} {\bibinfo {author} {\bibfnamefont {Y.}~\bibnamefont
  {Gu}},\ }\emph {\bibinfo {title} {Moments of Random Matrices and. Weingarten
  Functions}},\ \href {http://hdl.handle.net/1974/8241} {Master's thesis},\
  \bibinfo  {school} {Queen's University}, \bibinfo {address} {Ontario, Canada}
  (\bibinfo {year} {2013})\BibitemShut {NoStop}%
\bibitem [{\citenamefont {Collins}\ and\ \citenamefont
  {{\'{S}}niady}(2006)}]{Collins}%
  \BibitemOpen
  \bibfield  {author} {\bibinfo {author} {\bibfnamefont {B.}~\bibnamefont
  {Collins}}\ and\ \bibinfo {author} {\bibfnamefont {P.}~\bibnamefont
  {{\'{S}}niady}},\ }\bibfield  {title} {\enquote {\bibinfo {title}
  {Integration with respect to the {Haar} measure on unitary, orthogonal and
  symplectic group},}\ }\href {\doibase 10.1007/s00220-006-1554-3} {\bibfield
  {journal} {\bibinfo  {journal} {Commun. Math. Phys.}\ }\textbf {\bibinfo
  {volume} {264}},\ \bibinfo {pages} {773} (\bibinfo {year}
  {2006})}\BibitemShut {NoStop}%
\bibitem [{\citenamefont {Pucha\l{}a}\ and\ \citenamefont
  {Miszczak}(2017)}]{Puchala}%
  \BibitemOpen
  \bibfield  {author} {\bibinfo {author} {\bibfnamefont {Z.}~\bibnamefont
  {Pucha\l{}a}}\ and\ \bibinfo {author} {\bibfnamefont {J.~A.}\ \bibnamefont
  {Miszczak}},\ }\bibfield  {title} {\enquote {\bibinfo {title} {Symbolic
  integration with respect to the {H}aar measure on the unitary groups},}\
  }\href {\doibase 10.1515/bpasts-2017-0003} {\bibfield  {journal} {\bibinfo
  {journal} {Bull. Pol. Acad. Sci. Tech. Sci.}\ }\textbf {\bibinfo {volume}
  {65}},\ \bibinfo {pages} {21} (\bibinfo {year} {2017})}\BibitemShut {NoStop}%
\bibitem [{\citenamefont {Roberts}\ and\ \citenamefont
  {Yoshida}(2017)}]{chaosYoshida}%
  \BibitemOpen
  \bibfield  {author} {\bibinfo {author} {\bibfnamefont {D.~A.}\ \bibnamefont
  {Roberts}}\ and\ \bibinfo {author} {\bibfnamefont {B.}~\bibnamefont
  {Yoshida}},\ }\bibfield  {title} {\enquote {\bibinfo {title} {Chaos and
  complexity by design},}\ }\href {\doibase 10.1007/JHEP04(2017)121} {\bibfield
   {journal} {\bibinfo  {journal} {J. High Energy Phys.}\ }\textbf {\bibinfo
  {volume} {2017}},\ \bibinfo {pages} {121} (\bibinfo {year}
  {2017})}\BibitemShut {NoStop}%
\bibitem [{\citenamefont {Brand{\~a}o}\ \emph {et~al.}(2016)\citenamefont
  {Brand{\~a}o}, \citenamefont {Harrow},\ and\ \citenamefont
  {Horodecki}}]{Brandao2016}%
  \BibitemOpen
  \bibfield  {author} {\bibinfo {author} {\bibfnamefont {F.}~\bibnamefont
  {Brand{\~a}o}}, \bibinfo {author} {\bibfnamefont {A.~W.}\ \bibnamefont
  {Harrow}}, \ and\ \bibinfo {author} {\bibfnamefont {M.}~\bibnamefont
  {Horodecki}},\ }\bibfield  {title} {\enquote {\bibinfo {title} {Local random
  quantum circuits are approximate polynomial-designs},}\ }\href {\doibase
  10.1007/s00220-016-2706-8} {\bibfield  {journal} {\bibinfo  {journal}
  {Commun. Math. Phys.}\ }\textbf {\bibinfo {volume} {346}},\ \bibinfo {pages}
  {397} (\bibinfo {year} {2016})}\BibitemShut {NoStop}%
\bibitem [{\citenamefont {Gl{\"a}{\ss}le}(2013)}]{AlmostAllPureMME}%
  \BibitemOpen
  \bibfield  {author} {\bibinfo {author} {\bibfnamefont {T.}~\bibnamefont
  {Gl{\"a}{\ss}le}},\ }\href@noop {} {\enquote {\bibinfo {title} {Seminar 8 of
  selected topics in mathematical physics: Quantum information theory},}\
  }\bibinfo {howpublished} {\url{http://www.mpmueller.net/seminar/talk8.pdf}}
  (\bibinfo {year} {2013})\BibitemShut {NoStop}%
\bibitem [{\citenamefont {Lubkin}(1978)}]{Lubkin}%
  \BibitemOpen
  \bibfield  {author} {\bibinfo {author} {\bibfnamefont {E.}~\bibnamefont
  {Lubkin}},\ }\bibfield  {title} {\enquote {\bibinfo {title} {Entropy of an
  n-system from its correlation with a k-reservoir},}\ }\href {\doibase
  10.1063/1.523763} {\bibfield  {journal} {\bibinfo  {journal} {J. Math.
  Phys.}\ }\textbf {\bibinfo {volume} {19}},\ \bibinfo {pages} {1028} (\bibinfo
  {year} {1978})}\BibitemShut {NoStop}%
\bibitem [{\citenamefont {Page}(1993)}]{PageEntropy}%
  \BibitemOpen
  \bibfield  {author} {\bibinfo {author} {\bibfnamefont {D.~N.}\ \bibnamefont
  {Page}},\ }\bibfield  {title} {\enquote {\bibinfo {title} {Average entropy of
  a subsystem},}\ }\href {\doibase 10.1103/PhysRevLett.71.1291} {\bibfield
  {journal} {\bibinfo  {journal} {Phys. Rev. Lett.}\ }\textbf {\bibinfo
  {volume} {71}},\ \bibinfo {pages} {1291} (\bibinfo {year}
  {1993})}\BibitemShut {NoStop}%
\bibitem [{\citenamefont {Lloyd}\ and\ \citenamefont
  {Pagels}(1988)}]{LloydPagels}%
  \BibitemOpen
  \bibfield  {author} {\bibinfo {author} {\bibfnamefont {S.}~\bibnamefont
  {Lloyd}}\ and\ \bibinfo {author} {\bibfnamefont {H.}~\bibnamefont {Pagels}},\
  }\bibfield  {title} {\enquote {\bibinfo {title} {Complexity as thermodynamic
  depth},}\ }\href {\doibase 10.1016/0003-4916(88)90094-2} {\bibfield
  {journal} {\bibinfo  {journal} {Ann. Phys. New York}\ }\textbf {\bibinfo
  {volume} {188}},\ \bibinfo {pages} {186 } (\bibinfo {year}
  {1988})}\BibitemShut {NoStop}%
\bibitem [{\citenamefont {Scott}\ and\ \citenamefont
  {Caves}(2003)}]{ScottCaves}%
  \BibitemOpen
  \bibfield  {author} {\bibinfo {author} {\bibfnamefont {A.~J.}\ \bibnamefont
  {Scott}}\ and\ \bibinfo {author} {\bibfnamefont {C.~M.}\ \bibnamefont
  {Caves}},\ }\bibfield  {title} {\enquote {\bibinfo {title} {Entangling power
  of the quantum baker's map},}\ }\href {\doibase 10.1088/0305-4470/36/36/308}
  {\bibfield  {journal} {\bibinfo  {journal} {J. Phys. A: Math. Gen.}\ }\textbf
  {\bibinfo {volume} {36}},\ \bibinfo {pages} {9553} (\bibinfo {year}
  {2003})}\BibitemShut {NoStop}%
\bibitem [{\citenamefont {Giraud}(2007)}]{Giraud}%
  \BibitemOpen
  \bibfield  {author} {\bibinfo {author} {\bibfnamefont {O.}~\bibnamefont
  {Giraud}},\ }\bibfield  {title} {\enquote {\bibinfo {title} {Purity
  distribution for bipartite random pure states},}\ }\href {\doibase
  10.1088/1751-8113/40/49/f03} {\bibfield  {journal} {\bibinfo  {journal} {J.
  Phys. A: Math. Theor.}\ }\textbf {\bibinfo {volume} {40}},\ \bibinfo {pages}
  {F1053} (\bibinfo {year} {2007})}\BibitemShut {NoStop}%
\bibitem [{\citenamefont {Pasquale}\ \emph {et~al.}(2012)\citenamefont
  {Pasquale}, \citenamefont {Facchi}, \citenamefont {Giovannetti},
  \citenamefont {Parisi}, \citenamefont {Pascazio},\ and\ \citenamefont
  {Scardicchio}}]{Pasquale}%
  \BibitemOpen
  \bibfield  {author} {\bibinfo {author} {\bibfnamefont {A.~D.}\ \bibnamefont
  {Pasquale}}, \bibinfo {author} {\bibfnamefont {P.}~\bibnamefont {Facchi}},
  \bibinfo {author} {\bibfnamefont {V.}~\bibnamefont {Giovannetti}}, \bibinfo
  {author} {\bibfnamefont {G.}~\bibnamefont {Parisi}}, \bibinfo {author}
  {\bibfnamefont {S.}~\bibnamefont {Pascazio}}, \ and\ \bibinfo {author}
  {\bibfnamefont {A.}~\bibnamefont {Scardicchio}},\ }\bibfield  {title}
  {\enquote {\bibinfo {title} {Statistical distribution of the local purity in
  a large quantum system},}\ }\href {\doibase 10.1088/1751-8113/45/1/015308}
  {\bibfield  {journal} {\bibinfo  {journal} {J. Phys. A: Math. Theor.}\
  }\textbf {\bibinfo {volume} {45}},\ \bibinfo {pages} {015308} (\bibinfo
  {year} {2012})}\BibitemShut {NoStop}%
\bibitem [{\citenamefont {Mezzadri}(2007)}]{Mezzadri}%
  \BibitemOpen
  \bibfield  {author} {\bibinfo {author} {\bibfnamefont {F.}~\bibnamefont
  {Mezzadri}},\ }\bibfield  {title} {\enquote {\bibinfo {title} {How to
  generate random matrices from the classical compact groups},}\ }\href
  {https://www.ams.org/journals/notices/200705/fea-mezzadri-web.pdf} {\bibfield
   {journal} {\bibinfo  {journal} {Notices of the AMS}\ }\textbf {\bibinfo
  {volume} {54}},\ \bibinfo {pages} {592} (\bibinfo {year} {2007})}\BibitemShut
  {NoStop}%
\bibitem [{\citenamefont {Cotler}\ \emph {et~al.}(2017)\citenamefont {Cotler},
  \citenamefont {Hunter-Jones}, \citenamefont {Liu},\ and\ \citenamefont
  {Yoshida}}]{Cotler2017}%
  \BibitemOpen
  \bibfield  {author} {\bibinfo {author} {\bibfnamefont {J.}~\bibnamefont
  {Cotler}}, \bibinfo {author} {\bibfnamefont {N.}~\bibnamefont
  {Hunter-Jones}}, \bibinfo {author} {\bibfnamefont {J.}~\bibnamefont {Liu}}, \
  and\ \bibinfo {author} {\bibfnamefont {B.}~\bibnamefont {Yoshida}},\
  }\bibfield  {title} {\enquote {\bibinfo {title} {Chaos, complexity, and
  random matrices},}\ }\href {\doibase 10.1007/JHEP11(2017)048} {\bibfield
  {journal} {\bibinfo  {journal} {J. High Energy Phys.}\ }\textbf {\bibinfo
  {volume} {2017}},\ \bibinfo {pages} {48} (\bibinfo {year}
  {2017})}\BibitemShut {NoStop}%
\bibitem [{\citenamefont {Low}(2009)}]{Low}%
  \BibitemOpen
  \bibfield  {author} {\bibinfo {author} {\bibfnamefont {R.~A.}\ \bibnamefont
  {Low}},\ }\bibfield  {title} {\enquote {\bibinfo {title} {Large deviation
  bounds for k-designs},}\ }\href {\doibase 10.1098/rspa.2009.0232} {\bibfield
  {journal} {\bibinfo  {journal} {Proc. R. Soc. A}\ }\textbf {\bibinfo {volume}
  {465}},\ \bibinfo {pages} {3289} (\bibinfo {year} {2009})}\BibitemShut
  {NoStop}%
\bibitem [{\citenamefont {Nakata}\ \emph {et~al.}(2017)\citenamefont {Nakata},
  \citenamefont {Hirche}, \citenamefont {Koashi},\ and\ \citenamefont
  {Winter}}]{PhysRevX.7.021006}%
  \BibitemOpen
  \bibfield  {author} {\bibinfo {author} {\bibfnamefont {Y.}~\bibnamefont
  {Nakata}}, \bibinfo {author} {\bibfnamefont {C.}~\bibnamefont {Hirche}},
  \bibinfo {author} {\bibfnamefont {M.}~\bibnamefont {Koashi}}, \ and\ \bibinfo
  {author} {\bibfnamefont {A.}~\bibnamefont {Winter}},\ }\bibfield  {title}
  {\enquote {\bibinfo {title} {Efficient quantum pseudorandomness with nearly
  time-independent {H}amiltonian dynamics},}\ }\href {\doibase
  10.1103/PhysRevX.7.021006} {\bibfield  {journal} {\bibinfo  {journal} {Phys.
  Rev. X}\ }\textbf {\bibinfo {volume} {7}},\ \bibinfo {pages} {021006}
  (\bibinfo {year} {2017})}\BibitemShut {NoStop}%
\bibitem [{\citenamefont {Milz}\ \emph
  {et~al.}(2017{\natexlab{b}})\citenamefont {Milz}, \citenamefont {Sakuldee},
  \citenamefont {Pollock},\ and\ \citenamefont {Modi}}]{milz_kolmogorov_2017}%
  \BibitemOpen
  \bibfield  {author} {\bibinfo {author} {\bibfnamefont {S.}~\bibnamefont
  {Milz}}, \bibinfo {author} {\bibfnamefont {F.}~\bibnamefont {Sakuldee}},
  \bibinfo {author} {\bibfnamefont {F.~A.}\ \bibnamefont {Pollock}}, \ and\
  \bibinfo {author} {\bibfnamefont {K.}~\bibnamefont {Modi}},\ }\bibfield
  {title} {\enquote {\bibinfo {title} {Kolmogorov extension theorem for
  (quantum) causal modelling and general probabilistic theories},}\ }\href
  {https://arxiv.org/abs/1712.02589} {\bibfield  {journal} {\bibinfo  {journal}
  {arXiv:1712.02589}\ } (\bibinfo {year} {2017}{\natexlab{b}})}\BibitemShut
  {NoStop}%
\bibitem [{\citenamefont {Anza}\ \emph {et~al.}(2018)\citenamefont {Anza},
  \citenamefont {Gogolin},\ and\ \citenamefont
  {Huber}}]{PhysRevLett.120.150603}%
  \BibitemOpen
  \bibfield  {author} {\bibinfo {author} {\bibfnamefont {F.}~\bibnamefont
  {Anza}}, \bibinfo {author} {\bibfnamefont {C.}~\bibnamefont {Gogolin}}, \
  and\ \bibinfo {author} {\bibfnamefont {M.}~\bibnamefont {Huber}},\ }\bibfield
   {title} {\enquote {\bibinfo {title} {Eigenstate thermalization for
  degenerate observables},}\ }\href {\doibase 10.1103/PhysRevLett.120.150603}
  {\bibfield  {journal} {\bibinfo  {journal} {Phys. Rev. Lett.}\ }\textbf
  {\bibinfo {volume} {120}},\ \bibinfo {pages} {150603} (\bibinfo {year}
  {2018})}\BibitemShut {NoStop}%
\bibitem [{\citenamefont {Pollock}\ and\ \citenamefont
  {Modi}(2018)}]{arXiv:1704.06204}%
  \BibitemOpen
  \bibfield  {author} {\bibinfo {author} {\bibfnamefont {F.~A.}\ \bibnamefont
  {Pollock}}\ and\ \bibinfo {author} {\bibfnamefont {K.}~\bibnamefont {Modi}},\
  }\bibfield  {title} {\enquote {\bibinfo {title} {Tomographically
  reconstructed master equations for any open quantum dynamics},}\ }\href
  {\doibase 10.22331/q-2018-07-11-76} {\bibfield  {journal} {\bibinfo
  {journal} {Quantum}\ }\textbf {\bibinfo {volume} {2}},\ \bibinfo {pages} {76}
  (\bibinfo {year} {2018})}\BibitemShut {NoStop}%
\bibitem [{\citenamefont {Tamascelli}\ \emph {et~al.}(2018)\citenamefont
  {Tamascelli}, \citenamefont {Smirne}, \citenamefont {Huelga},\ and\
  \citenamefont {Plenio}}]{tamascelli2018}%
  \BibitemOpen
  \bibfield  {author} {\bibinfo {author} {\bibfnamefont {D.}~\bibnamefont
  {Tamascelli}}, \bibinfo {author} {\bibfnamefont {A.}~\bibnamefont {Smirne}},
  \bibinfo {author} {\bibfnamefont {S.~F.}\ \bibnamefont {Huelga}}, \ and\
  \bibinfo {author} {\bibfnamefont {M.~B.}\ \bibnamefont {Plenio}},\ }\bibfield
   {title} {\enquote {\bibinfo {title} {Nonperturbative treatment of
  non-{Markovian} dynamics of open quantum systems},}\ }\href {\doibase
  10.1103/PhysRevLett.120.030402} {\bibfield  {journal} {\bibinfo  {journal}
  {Phys. Rev. Lett.}\ }\textbf {\bibinfo {volume} {120}},\ \bibinfo {pages}
  {030402} (\bibinfo {year} {2018})}\BibitemShut {NoStop}%
\bibitem [{\citenamefont {Luchnikov}\ \emph {et~al.}(2019)\citenamefont
  {Luchnikov}, \citenamefont {Vintskevich}, \citenamefont {Ouerdane},\ and\
  \citenamefont {Filippov}}]{Luchnikov2018}%
  \BibitemOpen
  \bibfield  {author} {\bibinfo {author} {\bibfnamefont {I.~A.}\ \bibnamefont
  {Luchnikov}}, \bibinfo {author} {\bibfnamefont {S.~V.}\ \bibnamefont
  {Vintskevich}}, \bibinfo {author} {\bibfnamefont {H.}~\bibnamefont
  {Ouerdane}}, \ and\ \bibinfo {author} {\bibfnamefont {S.~N.}\ \bibnamefont
  {Filippov}},\ }\bibfield  {title} {\enquote {\bibinfo {title} {Simulation
  complexity of open quantum dynamics: Connection with tensor networks},}\
  }\href {\doibase 10.1103/PhysRevLett.122.160401} {\bibfield  {journal}
  {\bibinfo  {journal} {Phys. Rev. Lett.}\ }\textbf {\bibinfo {volume} {122}},\
  \bibinfo {pages} {160401} (\bibinfo {year} {2019})}\BibitemShut {NoStop}%
\bibitem [{\citenamefont {Arias}\ \emph {et~al.}(2002)\citenamefont {Arias},
  \citenamefont {Gheondea},\ and\ \citenamefont {Gudder}}]{fixedpoint}%
  \BibitemOpen
  \bibfield  {author} {\bibinfo {author} {\bibfnamefont {A.}~\bibnamefont
  {Arias}}, \bibinfo {author} {\bibfnamefont {A.}~\bibnamefont {Gheondea}}, \
  and\ \bibinfo {author} {\bibfnamefont {S.}~\bibnamefont {Gudder}},\
  }\bibfield  {title} {\enquote {\bibinfo {title} {Fixed points of quantum
  operations},}\ }\href {\doibase 10.1063/1.1519669} {\bibfield  {journal}
  {\bibinfo  {journal} {J. Math. Phys.}\ }\textbf {\bibinfo {volume} {43}},\
  \bibinfo {pages} {5872} (\bibinfo {year} {2002})}\BibitemShut {NoStop}%
\bibitem [{\citenamefont {Cramer}(2012)}]{Cramer}%
  \BibitemOpen
  \bibfield  {author} {\bibinfo {author} {\bibfnamefont {M.}~\bibnamefont
  {Cramer}},\ }\bibfield  {title} {\enquote {\bibinfo {title} {Thermalization
  under randomized local {Hamiltonians}},}\ }\href {\doibase
  10.1088/1367-2630/14/5/053051} {\bibfield  {journal} {\bibinfo  {journal}
  {New J. Phys.}\ }\textbf {\bibinfo {volume} {14}},\ \bibinfo {pages} {053051}
  (\bibinfo {year} {2012})}\BibitemShut {NoStop}%
\bibitem [{\citenamefont {Weingarten}(1978)}]{Weingarten}%
  \BibitemOpen
  \bibfield  {author} {\bibinfo {author} {\bibfnamefont {D.}~\bibnamefont
  {Weingarten}},\ }\bibfield  {title} {\enquote {\bibinfo {title} {Asymptotic
  behavior of group integrals in the limit of infinite rank},}\ }\href
  {\doibase 10.1063/1.523807} {\bibfield  {journal} {\bibinfo  {journal} {J.
  Math. Phys.}\ }\textbf {\bibinfo {volume} {19}},\ \bibinfo {pages} {999}
  (\bibinfo {year} {1978})}\BibitemShut {NoStop}%
\bibitem [{\citenamefont {Ginory}\ and\ \citenamefont {Kim}(2016)}]{IntHaar}%
  \BibitemOpen
  \bibfield  {author} {\bibinfo {author} {\bibfnamefont {A.}~\bibnamefont
  {Ginory}}\ and\ \bibinfo {author} {\bibfnamefont {J.}~\bibnamefont {Kim}},\
  }\bibfield  {title} {\enquote {\bibinfo {title} {Weingarten calculus and the
  {I}nt{H}aar package for integrals over compact matrix groups},}\ }\href
  {https://arxiv.org/abs/1612.07641} {\bibfield  {journal} {\bibinfo  {journal}
  {arXiv:1612.07641}\ } (\bibinfo {year} {2016})}\BibitemShut {NoStop}%
\bibitem [{\citenamefont {Taranto}\ \emph {et~al.}(2018)\citenamefont
  {Taranto}, \citenamefont {Modi},\ and\ \citenamefont {Pollock}}]{Philthy}%
  \BibitemOpen
  \bibfield  {author} {\bibinfo {author} {\bibfnamefont {P.}~\bibnamefont
  {Taranto}}, \bibinfo {author} {\bibfnamefont {K.}~\bibnamefont {Modi}}, \
  and\ \bibinfo {author} {\bibfnamefont {F.~A.}\ \bibnamefont {Pollock}},\
  }\bibfield  {title} {\enquote {\bibinfo {title} {Emergence of a fluctuation
  relation for heat in nonequilibrium {Landauer} processes},}\ }\href {\doibase
  10.1103/PhysRevE.97.052111} {\bibfield  {journal} {\bibinfo  {journal} {Phys.
  Rev. E}\ }\textbf {\bibinfo {volume} {97}},\ \bibinfo {pages} {052111}
  (\bibinfo {year} {2018})}\BibitemShut {NoStop}%
\bibitem [{\citenamefont {Epstein}\ and\ \citenamefont
  {Whaley}(2017)}]{qspeedLipsch}%
  \BibitemOpen
  \bibfield  {author} {\bibinfo {author} {\bibfnamefont {J.~M.}\ \bibnamefont
  {Epstein}}\ and\ \bibinfo {author} {\bibfnamefont {K.~B.}\ \bibnamefont
  {Whaley}},\ }\bibfield  {title} {\enquote {\bibinfo {title} {Quantum speed
  limits for quantum-information-processing tasks},}\ }\href {\doibase
  10.1103/PhysRevA.95.042314} {\bibfield  {journal} {\bibinfo  {journal} {Phys.
  Rev. A}\ }\textbf {\bibinfo {volume} {95}},\ \bibinfo {pages} {042314}
  (\bibinfo {year} {2017})}\BibitemShut {NoStop}%
\bibitem [{\citenamefont {Deza}\ and\ \citenamefont
  {Deza}(2009)}]{deza2009encyclopedia}%
  \BibitemOpen
  \bibfield  {author} {\bibinfo {author} {\bibfnamefont {M.}~\bibnamefont
  {Deza}}\ and\ \bibinfo {author} {\bibfnamefont {E.}~\bibnamefont {Deza}},\
  }\href {\doibase 10.1007/978-3-642-00234-2} {\emph {\bibinfo {title}
  {Encyclopedia of Distances}}},\ Encyclopedia of Distances\ (\bibinfo
  {publisher} {Springer Berlin Heidelberg},\ \bibinfo {year}
  {2009})\BibitemShut {NoStop}%
\bibitem [{\citenamefont {Zhang}\ and\ \citenamefont
  {Xiang}(2017)}]{ZhangXiang}%
  \BibitemOpen
  \bibfield  {author} {\bibinfo {author} {\bibfnamefont {L.}~\bibnamefont
  {Zhang}}\ and\ \bibinfo {author} {\bibfnamefont {H.}~\bibnamefont {Xiang}},\
  }\bibfield  {title} {\enquote {\bibinfo {title} {Average entropy of a
  subsystem over a global unitary orbit of a mixed bipartite state},}\ }\href
  {\doibase 10.1007/s11128-017-1570-6} {\bibfield  {journal} {\bibinfo
  {journal} {Quantum Inf. Process.}\ }\textbf {\bibinfo {volume} {16}},\
  \bibinfo {pages} {112} (\bibinfo {year} {2017})}\BibitemShut {NoStop}%
\end{thebibliography}
\end{document}